\documentclass[reprint,superscriptaddress,twocolumn,amsmath,amssymb,]{revtex4-2}
\pdfoutput=1
\usepackage{footmisc}
\usepackage{xcolor}
\usepackage{graphicx}
\usepackage{dcolumn}
\usepackage{bm}
\usepackage{romannum}
\usepackage{siunitx}
\usepackage{epstopdf}
\usepackage{float}
\usepackage{titlesec}
\titleformat{\section}{\raggedright\fontsize{12.5}{25}\bfseries}{\arabic{section}.}{1em}{}

\DeclareUnicodeCharacter{0301}{\'a}
\DeclareUnicodeCharacter{00D6}{\"o}
\DeclareUnicodeCharacter{00C4}{\"a}
\DeclareUnicodeCharacter{00FC}{\"u}
\DeclareUnicodeCharacter{00DF}{\ss}

\usepackage{hyperref}
\usepackage[normalem]{ulem}

\begin{document}

\pagenumbering{arabic}

\title{\fontsize{22}{19}\selectfont  Nonequilibrium dynamics of $\alpha$-RuCl$_{3}$ - a time-resolved magneto-optical spectroscopy study}

\author{Julian Wagner}

\author{Anuja Sahasrabudhe}

\author{Rolf Versteeg}
\affiliation{Universität zu Köln, II. Physikalisches Institut, Zülpicher Straße 77, Köln D-50937, Germany}

\author{Zhe Wang}
\affiliation{Department of Physics, TU Dortmund University, Otto-Hahn-Str. 4, 44227 Dortmund, Germany}

\author{Vladimir Tsurkan}
\affiliation{Experimental Physics V, Center for Electronic Correlations and Magnetism, University of Augsburg, 86135 Augsburg, Germany}
\affiliation{ Institute of Applied Physics, MD 2028, Chisinau, Republic of Moldova}

\author{Alois Loidl}
\affiliation{Experimental Physics V, Center for Electronic Correlations and Magnetism, University of Augsburg, 86135 Augsburg, Germany}

\author{Hamoon Hedayat}
\email{hedayat@ph2.uni-koeln.de}
\author{Paul H. M. van Loosdrecht}
\email{pvl@ph2.uni-koeln.de}
\affiliation{Universität zu Köln, II. Physikalisches Institut, Zülpicher Straße 77, Köln D-50937, Germany}

\begin{abstract}
We present time-resolved magneto-optical spectroscopy on the magnetic Mott-Hubbard-insulating Kitaev spin liquid candidate $\alpha$-RuCl$_3$ to investigate the nonequilibrium dynamics of its antiferromagnetically ordered zigzag groundstate after photoexcitation. A systematic study of the transient magnetic linear dichroism under different experimental conditions (temperature, external magnetic field, photoexcitation density) gives direct access to the dynamical interplay of charge excitations with the zigzag ordered state on ultrashort time scales. We observe a rather slow initial demagnetization (few to 10s of ps) followed by a long-lived non-thermal antiferromagnetic spin-disordered state (100$-$1000s of ps), which can be understood in terms of holons and doublons disordering the antiferromagnetic background after photoexcitation. Varying temperature and fluence in the presence of an external magnetic field reveals two distinct photoinduced dynamics associated with the zigzag and quantum paramagnetic disordered phases. The photo-induced non-thermal spin-disordered state shows universal compressed-exponential recovery dynamics related to the growth and propagation of zigzag domains on nanosecond time scales, which is interpreted within the framework of the Fatuzzo-Labrune model for magnetization reversal. The study of nonequilibrium states in strongly correlated materials is a relatively unexplored topic, but our results are expected to be extendable to a large class of Mott-Hubbard insulator materials with strong spin-orbit coupling.
\end{abstract}

\maketitle
\section{Introduction}

\begin{figure*}
\includegraphics[scale=0.50]{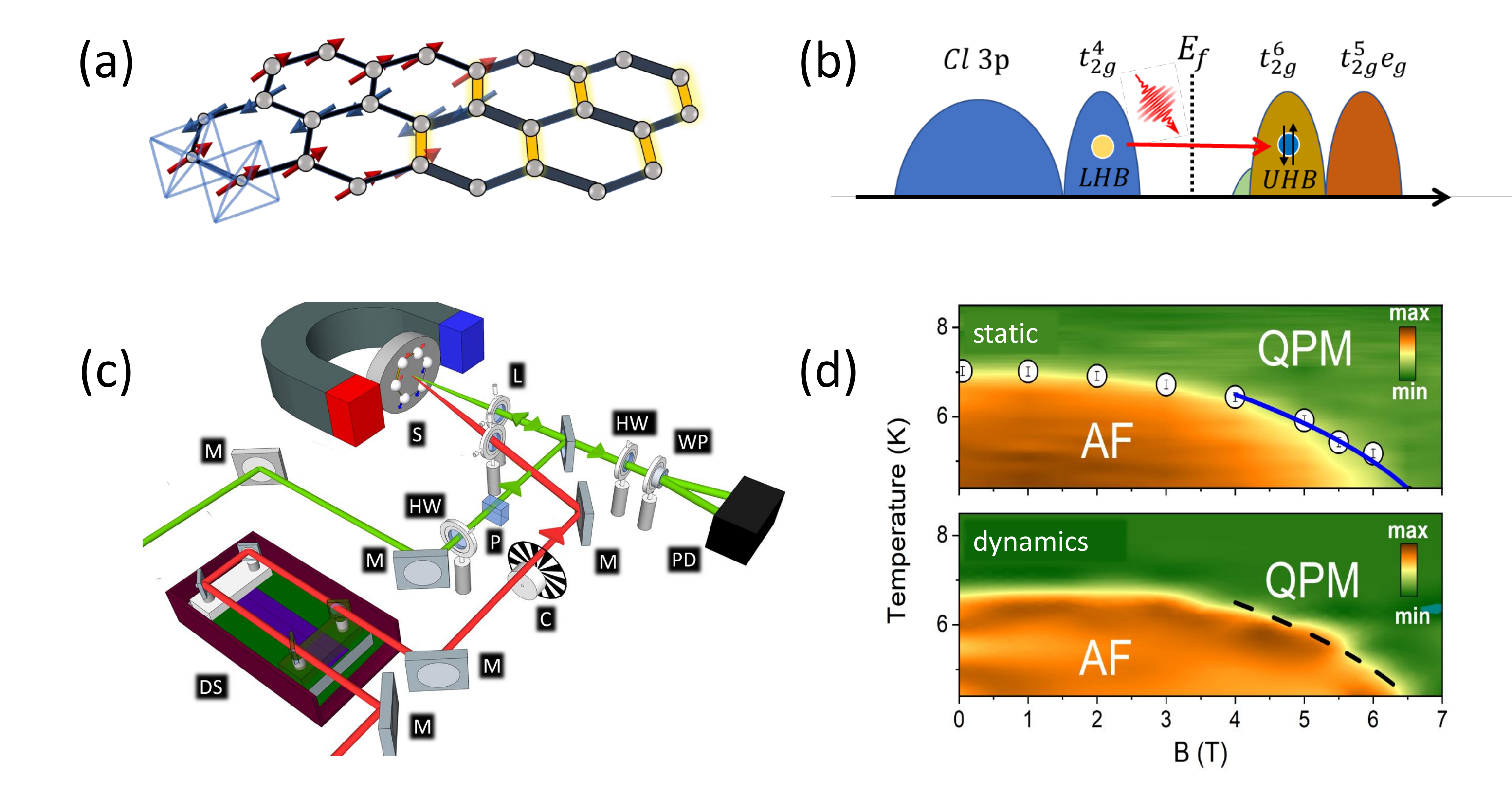}
\caption{{\bfseries{Time-resolved magneto-optical experiment}}. (a) Crystal structure and zigzag spin orientation. The color contrast of yellow and dark blue bonds indicates small inequivalence in one of the Ru-Ru bond lengths \cite{cao2016low}. (b)  Schematical electronic structure and photo-excitation process across the Mott-Hubbard gap. (c) Experimental setup (HW: half-wave plate, P: polarizer, DS: delay stage, C: chopper, M: (beam-splitting) mirror, L: lens, S: sample, WP: Wollaston prism, PD: photo-diode). (d) In-plane magnetic field and temperature phase diagram of $\alpha$-RuCl$_3$ constructed from the static change in rotation $\theta_\text{MLD}$ obtained by iso-magnetic MLD measurements with a fluence of $0.6$ W/cm$^2$ (upper panel) and extracted from the time-integrated transient MLD response $\Delta \theta_\text{MLD}(t)$ for an applied pump fluence of \SI{3}{\micro\joule/\cm^2} (lower panel). White circles in the upper panel indicate the Néel temperature derived from the derivative of the rotation w.r.t. the temperature \cite{wagnerstaticMLD}. The color code displays the amplitude of the rotation from 0 to 70 mdeg (upper panel) and on a relative scale in the lower panel.}
\label{Fig1}
\end{figure*}

 To probe material properties using ultrafast laser pulses is a fascinating promise of modern experimental condensed matter physics, which has been shown to be very powerful in accessing new transient phases, which are nonexistent under thermal equilibrium conditions \cite{de2021nonthermal,li2018theory}. A prime example is the ultrafast manipulation of magnetic order \cite{kalashnikova2015ultrafast,kirilyuk2010ultrafast}. 
 Especially, the understanding of the underlying dynamical processes, associated time scales and how magnetic order can be controlled is at the forefront of non-equilibrium condensed matter research. Of particular interest are Mott-insulators, which exhibit strong electron-electron interactions and significant correlations between electronic and magnetic degrees of freedom giving rise to exotic properties and groundstates \cite{jackeli2009mott}. An interesting case is found in the magnetic Mott-Hubbard-insulator $\alpha$-RuCl$_3$, which is believed to be a prime candidate to realize the so-far elusive Kitaev quantum spin liquid ground state, resulting from the presence of strong bond-dependent anisotropic coupling among spins accompanied by quantum fluctuations \cite{savary2016quantum,Kitaev20062}. The trihalide $\alpha$-RuCl$_3$ is a layered material, in which edge-sharing RuCl$_6$ octahedra with Ru$^{3+}$ ions in the center form a two-dimensional honeycomb structure (c.f. Fig. \ref{Fig1}(a)). The partially filled $t_{2g}$ orbitals are coupled via nearly $90^\circ$ superexchange paths \cite{glamazda2017relation}. The $4d^5$ electronic configuration of Ru is in a low-spin state with $S =1/2$, giving rise to $J_{eff} = 1/2$ pseudospins due to the presence of the octahedral crystal-field splitting and large spin-orbit coupling \cite{loidl2021proximate,trebst2017kitaev}. The on-site intercation $U$ opens a gap inside the $J_{eff}=1/2$ band, creating the upper Hubbard band (UHB) and lower Hubbard band (LHB), respectively. These properties are essential prerequisites to realize the Kitaev model in condensed matter systems. Thus, significant efforts have focused on the rich equilibrium phase diagram to find experimental fingerprints of this highly-entangled topological state of matter, which is characterized by spin-flip excitations fractionalizing into itinerant Majorana fermions and emergent gauge fields \cite{kasahara2018majorana,wen2019experimental}. 
 
 However, experimentally it has been found that $\alpha$-RuCl$_3$ establishes long-range antiferromagnetic zigzag (AFM) order below the Néel temperature of $\approx 7$ K and in-plane applied magnetic fields $\lesssim 7$ T, indicating the presence of significant non-Kitaev interactions \cite{fletcher1967x,sears2020ferromagnetic}. Above the Néel temperature the zigzag magnetic order is destroyed and $\alpha$-RuCl$_3$ exhibits a quantum paramagnetic phase (QPM) characterized by strong quantum fluctuations. We have investigated the zigzag phase in one of our recent studies, where we report on the anisotropic magneto-optical response under equilibrium conditions and show that magneto-optical spectroscopy is a versatile tool to study the antiferromagnetically zigzag ordered ground state by purely optical means \cite{wagnerstaticMLD}. 
 Currently, the discussion on $\alpha$-RuCl$_3$ equilibrium physics is still highly controversial, while there are experimental and theoretical arguments supporting that the magnetically ordered ground state of $\alpha$-RuCl$_3$ is unstable under external perturbations and metastable states are likely to occur under nonequilibrium conditions \cite{versteeg2020nonequilibrium,suzuki2021proximate,janssen2019heisenberg}. 
 
Further, besides being a spin liquid candidate material, the presence of Mott-Hubbard physics accompanied by strong spin-orbit coupling and the role of excitons in photoexcited $\alpha$-RuCl$_3$ makes it enormously complex with unique photoinduced dynamics providing an opportunity to investigate many-body  interactions. Understanding dynamical excitonic correlations and their interaction with a magnetically ordered background is a fundamental scientific question  \cite{bittner2020photoenhanced}. Optical spectroscopy and ellipsometry revealed a narrow line shape and asymmetry of the peak in the optical conductivity at around 1.1 eV associated with the optical gap, which suggest strong electron-hole interaction, and a small shoulder extending into the Mott gap indicating the presence of an excitonic bound state \cite{sandilands2016spin,sandilands2016optical}. Further, the excitonic fingerprint shows remarkable temperature dependence in equilibrium, indicating a coupling between high-energy excitations and the magnetic structure \cite{sandilands2016optical}. 

Here, open questions regarding the interaction of Mott-Hubbard excitons with other degrees of freedom and associated time scales arise. In our previous time-resolved magneto-optical spectroscopy study on the spin-liquid candidate $\alpha$-RuCl$_3$, we showed that the transient magnetic linear dichroism can be used to study the optically induced quench of antiferromagnetic order followed by a transient spin disordered state \cite{versteeg2020nonequilibrium}. It is known for decades that ultrafast laser pulses can be used to induce ultrafast demagnetization \cite{beaurepaire1996ultrafast,carpene2015ultrafast}. The demagnetization dynamics has been studied extensively and described successfully within different frameworks, e.g. the well-established phenomenological multi-temperature model \cite{sirica2021disentangling}, a time-dependent Ginzburg-Landau
theory approach \cite{versteeg2020nonequilibrium}, a combination of both models \cite{lovinger2020influence} or another recently developed general theory of out-of-equilibrium order parameter correlations \cite{dolgirev2020self}. Still, the microscopic interactions leading to magnetization dynamics in highly correlated materials has to be better understood. Especially the ultrafast melting of antiferromagnetic order in Mott-Insulators has been argued to be fundamentally different from metals and common band insulators due to the presence of enhanced electron-electron interactions. The (de)magnetization dynamics are believed to be mainly governed by a redistribution of charge carriers among neighboring lattice sites, leading to a fast generation of doubly occupied and empty sites, called \textit{doublons} and \textit{holons}. The coupling between charge excitations and the spin order has been argued to be the dominant mechanism leading to ultrafast spin dynamics \cite{afanasiev2019ultrafast}. Specifically, the motion of holons and doublons can create trails of defects in the antiferromagnetic background, transferring energy from the hot photocarriers to the antiferromagnetic order when these quasiparticles hop around. Local spin perturbations are then dispersed through magnons into the whole system. The recombination of holon-doublon pairs has been argued to be mediated by multimagnon emission, which provides an efficient electronic demagnetization mechanism to transfer the excess energy to the system \cite{lenarvcivc2013ultrafast,versteeg2020nonequilibrium}. This raises several questions, e.g., of how an applied magnetic field affects the dynamics of photo-excited holons and doublons. Further, it is important to understand whether the origin of the photo-induced dynamics is thermal or non-thermal and what are the potential mechanisms driving the recovery dynamics. In this regard, a more comprehensive study is needed to unravel the underlying physics.\\
To investigate the dynamics of the coupling between the charge and magnetic sector in $\alpha$-RuCl$_3$, we performed time-resolved optical pump-probe experiments, in which the system is excited by a pump laser pulse over the Mott-Hubbard gap $\Delta_\text{MH} \approx 1.1$~eV \cite{sandilands2016optical} and the transient magnetization dynamics are probed with time-delayed weaker probe laser pulses. Pumping $\alpha$-RuCl$_3$ with 1.55 eV photons, which have an excess energy over the MH gap, leads to multiple hopping processes via the impact ionization relaxation mechanism \cite{yamakawa2017mott} (c.f. Fig. \ref{Fig1}(b)).
The pump photons induce intersite $d$-$d$ charge transfer transitions resulting in doubly occupied and non-occupied sites creating \textit{multiple} spinless quasiparticles, the  doublons and holons. A recent time-resolved two-photon photoemission spectroscopy (2PPES) study reported the condensation of these quasiparticles forming Mott-Hubbard excitons within a few picoseconds after photoexcitation \cite{nevola2021timescales}, which are expected to interfere with the zigzag ordered antiferromagnetic background.\\ 
In this report, we systematically investigate the magnetization dynamics and the subsequent recovery processes from fs to ns time scales by transient magnetic linear dichroism measurements, which have been performed for different applied pump fluences varying by more than one order of magnitude, at different temperatures below and above the Néel temperature $T_N\approx 7$ K and in external magnetic fields strengths from $0-\pm7$ T applied along the $ab$ honeycomb-planes. Our experiments suggest that the presence of photo-induced holons and doublons gives rise to a transient long-lived spin disordered state, which cannot be reached by just tuning the temperature or magnetic field.

\section{Methods}

For the study of the dynamics of the antiferromagnetically ordered zigzag ground state in $\alpha$-RuCl$_3$, we apply a table-top pump-probe spectroscopy technique making use of the magnetic linear dichroism (MLD) effect in reflection geometry. Magnetic linear dichroism is a quadratic magneto-optical effect, in which a rotation of the light polarization (or a change of its ellipticity) occurs due to a difference of the complex index of refraction for normal incidence light with a polarization plane oriented parallel and perpendicular to the antiferromagnetic Néel vector, respectively \cite{ferre1984linear,mak2019probing}. The magnetic linear dichroism in antiferromagnets is sensitive to both the size and the orientation of the Néel vector, i.e. the order parameter. It has been used previously to determine the orientation of the Néel vector \cite{saidl2017optical}, to study zigzag antiferromagnetically ordered phases in equilibrium \cite{wagnerstaticMLD,zhang2021observation} and to track ultrafast dynamics of critical phenomena in antiferromagnetically ordered systems near a magnetic phase transition \cite{versteeg2020nonequilibrium,zhang2021spin}. 

For the study of the magneto-optical response the $\alpha$-RuCl$_3$ samples grown by vacuum sublimation \cite{sahasrabudhe2020high} were placed in a helium-cooled cryostat (Oxford Spectromag) with temperatures down to 2.2 K inside the coils of a superconducting magnet with magnetic field strengths up to $\pm7$~T. Characterization of the sample was done applying a static magnetic linear dichroism experiment \cite{wagnerstaticMLD}, which revealed a Néel temperature of $\approx 7$ K indicating good sample quality \cite{cao2016low,mi2021stacking}. Fig.~\ref{Fig1}(c) illustrates the experimental setup. The magnetic field was applied within the crystallographic $ab$-plane, i.e. within the honeycomb layers. The polarization of the incident probe light was rotated by a half waveplate and set to the polarization orientation setting, which gave the maximum signal in zero magnetic field. The pump beam passes through a computer-controlled delay line and is mechanically chopped at a frequency of $680$ Hz. The pump beams polarization is set cross-polarized to the probe beam in order to minimize potential pump beam scattering to reach the detector. The measurements of the magneto-optical response were carried out in the so-called Voigt geometry \cite{tesavrova2014systematic} at near-normal incidence, such that the light wave propagation vector $\bf{k}$ of the probe laser pulses was perpendicular to the honeycomb layer $ab$ planes ($\bf{k}\perp ab$) and magnetic field vector $\bf{B}_{ab}$ (see Fig.~\ref{Fig1}(a)). The time-resolved magneto-optical experiment was performed using 800 nm ($\approx 1.55$ eV) pump pulses with a temporal width of 40 fs, and probe pulses of 512 nm ($\approx 2.42$ eV) with a temporal width of 250 fs, generated with a laser system based on a \textsc{LightConversion Pharos} equiped with \textsc{LightConversion} optical parametric amplifiers as core components. The penetration depth of the pump is larger than for the probe, which helps to probe a homogeneously excited area \cite{versteeg2020nonequilibrium}. The pump and probe beam were focused down to a radius of \SI{40}{\micro\metre} and \SI{25}{\micro\metre}, respectively. The repetition rate of the amplified laser system was set to $f= 30$ kHz in order to ensure that the system can relax back to the ground state between consecutive pulses and photoinduced thermal heating is reduced. Detection of the transient polarization rotation $\Delta \theta(t)$ was done using a balanced-detection configuration \cite{prasankumar2016optical} consisting of a $\lambda/2$ wave plate, a Wollaston prism, and a balanced photodiode. The signal from the photodiodes was sent through a pre-amplifier before reading out the balanced signal by a \textsc{Zurich Instruments HF2LI} lock-in amplifier.

\begin{figure*}
\includegraphics[scale=0.5]{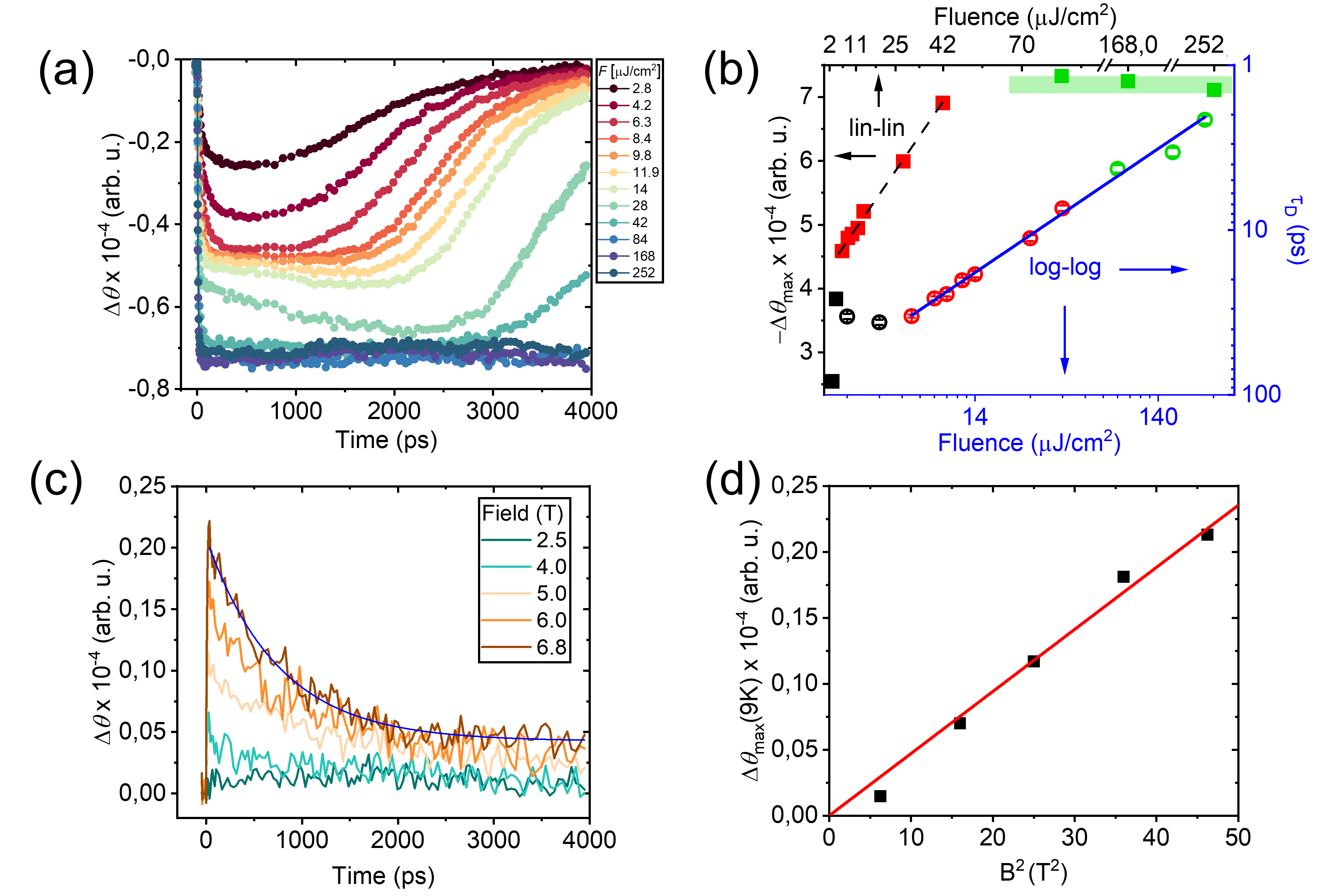}
\caption{{\bfseries{Comparison of the magneto-optical response in the zigzag AFM and QPM phase}}. (a) Fluence dependence of $\Delta \theta_\text{MLD}(t)$ in zero magnetic field at $3$ K for applied pump fluences varying between $3$ to \SI{250}{\micro\joule/\cm^2}. For applied fluences above \SI{70}{\micro\joule/\cm^2}, the maximum reached $\Delta \theta_\text{MLD}$ is constant. (b) Maximum photo-induced polarization rotation $\Delta \theta_\text{MLD}$ (squares, black axes) on a lin-lin scale and extracted demagnetization time constants $\tau_\text{D}$ (open circles, blue axes) on a log-log scale. A fit to extract power law scaling of $\tau_D$ is shown by the blue solid line. The color code of the symbols indicates the distinct fluence regimes: black, red and green represent the low, intermediate and high fluence regimes, respectively. Maximum demagnetization is indicated by the greenish background. (c) Field dependence of $\theta_\text{MLD}(t)$ at an applied fluence of \SI{80}{\micro\joule/\cm^2} in the quantum paramagnetic phase at $9$ K. The blue solid line is a single exponential fit with an offset. (d) Maximum amplitude of $\Delta \theta_\text{MLD}$ at $9$ K for an applied pump fluence of \SI{80}{\micro\joule/\cm^2} as a function of applied magnetic field, which scales linearly with $B^2$ as expected for second-order magnetic linear dichroism in the field-polarized QPM state.}
\label{Fig2}
\end{figure*}
\section{Results and discussion}

First, we show that changes in the magnetic linear dichroism $\Delta \theta_\text{MLD}$ are related to the magnetic order and probe the zigzag antiferromagnetic phase. The top panel of Fig.~\ref{Fig1}(d) shows the phase diagram as a function of temperature $T$ and in-plane applied magnetic field $B$ obtained under equilibrium conditions (same data as in \cite{wagnerstaticMLD}). The variation of $ \theta_\text{MLD}$ for different temperatures and fields reconstructs the already observed $B-T$ phase diagram of $\alpha$-RuCl$_3$ \cite{ma2018recent}. We note that the absolute value of $\theta_\text{MLD}$ varies by the formation of three equal zigzag domains depending on their angle with respect to the polarization orientation of the incident probe beam, such that it is non-trivial to extract the absolute change in rotation \cite{wagnerstaticMLD,saidl2017optical}. To check if the pump induced transient $\Delta \theta_\text{MLD}(t)$ is also directly related to the zigzag order, we measured $\Delta \theta_\text{MLD}(t)$ after photoexcitation and plotted the time-integrated signals as a function of equilibrium temperatures and fields in a false-color plot (see Fig.~\ref{Fig1}(b) lower panel). The fact that both from the transient $\Delta \theta_\text{MLD}(t)$ and the static $\theta_\text{MLD}$ the $B-T$ phase diagram can be reconstructed demonstrates the sensitivity of our approach to the zigzag antiferromagnetic order.

The fluence-dependent measurements of the induced change in the magnetic linear dichroism, $\Delta \theta_\text{MLD}(t)$, at a temperature of 3 K, i.e., deep in the zigzag ordered phase, are shown in Fig.~\ref{Fig2}(a). 
For small pump fluences, up to \SI{7}{\micro\joule/\cm^2} the transient magneto-optical response $\Delta \theta_\text{MLD}(t)$ reaches its maximum on a timescale of $\sim 400$ ps and then relaxes back to its pre-pump value within the accessible time-window of 4 ns. However, for increasing pump fluences the changes in $\Delta \theta_\text{MLD}(t)$ get more pronounced and a ns long-lived (metastable) state with plateauish, non-decaying behavior is observed. No further increase of $|\Delta \theta_\text{MLD}|$ is observed above a pump fluence of \SI{70}{\micro\joule/\cm^2}. We point out that while reducing the fluence again, all measurements are reproducible indicating no photo-induced sample damage. Further, the system can relax back to the equilibrium state between consecutive pulses as we do not observe any residual background, not even with the highest applied fluence, at negative time delays.

Fig.~\ref{Fig2}(b) shows the fluence dependence of the amplitude $\Delta \theta_\text{MLD}(t)$ at a time delay of 500 ps as well as the demagnetization time constants $\tau_{D}$ associated with the initial changes in the magneto-optical response after pump excitation. The fluence dependence of the amplitude $\Delta \theta_\text{MLD}(t)$ indicates \textit{three} distinct fluence regimes. In the low fluence regime (LFR) there is an initial sharp increase in $\Delta \theta_\text{MLD}(t)$ occurring for applied pump fluences up to \SI{7}{\micro\joule/\cm^2} (black symbols), followed by an intermediate rise between $10$ and \SI{50}{\micro\joule/\cm^2} (red symbols) called intermediate fluence regime (IFR). We observe saturation in the high fluence regime (HFR) above \SI{70}{\micro\joule/\cm^2} (green symbols), i.e., no further demagnetization occurs. In contrast, the demagnetization time constant $\tau_{D}$ extracted from single-exponential fits to the initial dynamics shows only two regimes. First, it is almost constant $\sim 30$ ps for fluences up to \SI{7}{\micro\joule/\cm^2}. 
Above this threshold fluence $\tau_{D}$ decreases non-linearly for increasing pump fluences $F$ and follows a clear power-law scaling $\propto F^{-0.7}$ (blue line in Fig.~\ref{Fig2}(b)). The fluence dependent behavior of $\Delta \theta_\text{MLD}(t)$ (amplitude of the demagnetization) and $\tau_{D}$ (risetime of the demagnetization) provide essential information on the photoexcited state. Together with other experimental observations presented later in this paper and other recently reported results \cite{versteeg2020nonequilibrium}, we will show that all three regimes are non-thermal. In the LFR, the ultrafastly created holons and doublons have already formed a bound pair. These excitons show an excitation density independent recombination process, which is indicated by the constant demagnetization time $\tau_D$ associated with the recombination via magnon emission \cite{lenarvcivc2013ultrafast}. If the holons and doublons would not be already bound, the recombination process would depend on the probability to encounter the oppositely charged particle that causes a non-exponential and excitation density dependent recombination process \cite{lenarvcivc2014charge,lenarvcivc2015nonequilibrium}. 
In the IFR, the excess amount of excited carriers gives rise to the screening effect, which causes a faster breaking of initially formed excitons into holons and doublons, which experimentally scales with the power-law $\propto F^{-0.7}$. The antiferromagnetic system is completely disordered in the highest excitation regime, such that no further change in $\Delta \theta_\text{MLD}(t)$) is observed, but the screening effect is continuously enhanced (decrease in $\tau_{D}$), as expected. The mechanism within the LER is conceivable, as already observed in similar Mott-Hubbard systems. At low excitation densities, the holons and doublons form bound pairs, a Mott-Hubbard exciton, (also reported for $\alpha$-RuCl$_3$ in \cite{nevola2021timescales}) which then recombine within 10s of ps, while their excess energy is dispersed through magnons and potentially other bosonic exciations into the whole system. The clearly different non-linear fluence dependence of the initial demagnetization dynamics within the IFR thus can be understood in terms of an increasing unbound holon-doublon density perturbing the zigzag order. The experimental observations indicate that the dominating demagnetization mechanism in the LFR and IFR seem to be different. For sufficiently small excitation densities holon-doublon pair recombination via magnon emission appears to dominate, the hopping of unbound holons and doublons in the IFR effectively perturbs the long-range zigzag order. 
Theory predicts that in the strong excitation regime indeed a larger fraction of the excitons separates into unbound doublons and holons \cite{bittner2020photoenhanced}. The most intriguing observation in the IFR is the analogous power-law scaling behavior ($\propto F^{-0.7}$) observed for the time scale of the gap collapse after photoexcitation in the excitonic insulator materials e.g. Ta$_2$NiSe$_5$ \cite{okazaki2018photo} and 1T-TiSe$_2$ \cite{rohwer2011collapse,hedayat2019excitonic} associated with a free carrier screening effect which is responsible that excitons break into electron and holes. However, the reported time scales are in the order of 100s of fs and different with the 10s to few ps timescale observed in our experiment. This could be related to the small bandgap in Ta$_2$NiSe$_5$ and 1T-TiSe$_2$, which is an order of magnitude smaller than for $\alpha$-RuCl$_3$. The characteristic build-up time for carrier screening in response to an ultrashort laser excitation is the plasma oscillation period, which scales with $n^{-0.5}$, where $n$ is the carrier density \cite{wagner2014ultrafast}. Since $\alpha$-RuCl$_3$ is a Mott-insulator and the carrier density in equilibrium expected to be small, the photo-excited carrier density should scale linearly with the applied fluence and hence a similar scaling behavior is expected. Indeed, the pump-induced changes in $\Delta\theta_\text{MLD}(t)$ in the IFR scales linearly with the applied fluence. Therefore, the excitation density dependent screening effect plays an important role in the dynamics in the IFR. In the HER, the photo-induced magenitude of the demagnetization is constant (in contrast with a saturating behavior of the risetime) suggesting a completely spin-disordered system. In the following, we elaborate in more detail on the interaction between holons, doublons and Mott-Hubbard excitons and the underlying magnetically ordered background. 

To prove experimentally that the presence of long-range zigzag antiferromagnetic correlations is crucial to observe the $\Delta \theta_\text{MLD}(t)$ dynamics in Fig.~\ref{Fig2} (a), we conducted measurements at 9 K in the disordered QPM phase of $\alpha$-RuCl$_3$ (see Fig.~\ref{Fig2} (c)). We point out that even at a high pump fluence of \SI{80}{\micro\joule/\cm^2} (cf. saturation of $\Delta \theta_\text{MLD}(t)$ at 3 K) transient changes in the polarization rotation $\Delta \theta_\text{MLD}(t)$ are very small in zero magnetic field in the QPM phase. This is the behavior, which is naturally expected, since in the QPM no long-range magnetic order exists and hence no magnetic linear dichroism is expected to be observed. However, by applying an external field, a transient change in $\Delta \theta_\text{MLD}(t)$ can be observed experimentally. The sign of $\Delta \theta_\text{MLD}(t)$ at 9 K is opposite to the one at 3 K. This is another evidence that the probed demagnetization dynamics originates from two distinct magnetic phases. The maximum photo-induced changes in $\Delta \theta_\text{MLD}(t)$ scales quadratically $\propto (\mu_0 H)^2$ in the externally applied field, which indicates that we are probing the magnetic linear dichroism in the field-aligned QPM phase with a finite net magnetization $\propto \chi H$ (cf. Fig.~\ref{Fig2}(d)). Moreover, the dynamics of $\Delta \theta_\text{MLD}(t)$ has a clearly different character than the dynamics in the AFM phase governed by a field-independent quick rise followed by a single exponential decay with a time constant of $\sim 800$ ps for all applied field strengths showing relaxation on ns time scales. As expected, there is no signature of a long-lived plateauish state. 

\begin{figure*}
\includegraphics[scale=0.55]{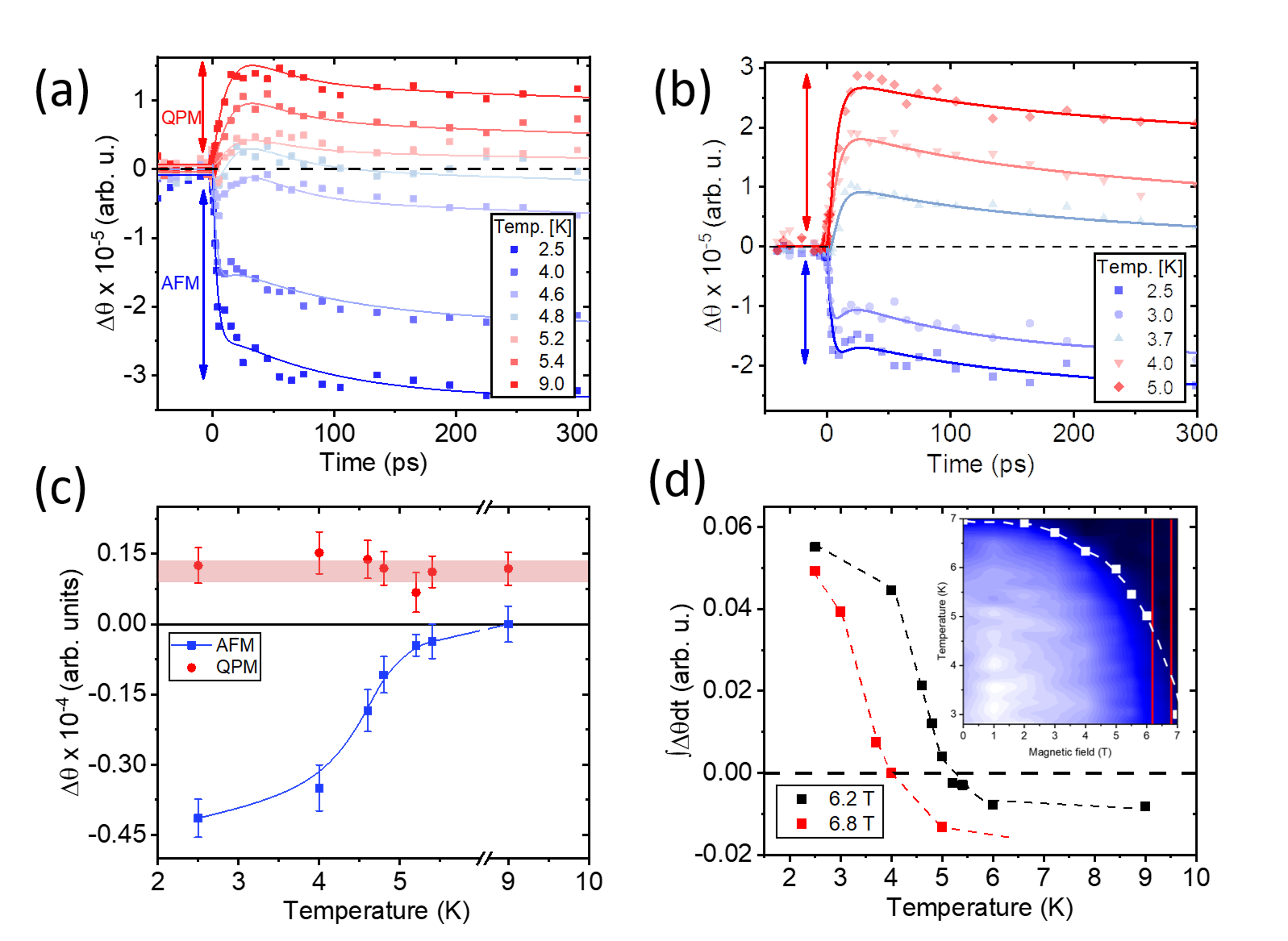}
\caption{{\bfseries{Coexistence of the QPM and AFM phase dynamics}}. Temperature dependence of $\theta_\text{MLD}(t)$ at an applied magnetic field of (a) $6.2$ T and (b) $6.8$ T for an applied pump fluence of $F=$\SI{80}{\micro\joule/\cm^2}. The solid lines display the fits described in the main text. (c) Extracted amplitudes of the two component exponential fit related to the zigzag AFM and the QPM phase for the fits derived from (a). The zigzag AFM component vanishes towards crossing the phase boundary, while the QPM dynamical component survives. (d) Changes of the integrated magneto-optical response as a function of temperature. The inset shows the $B-T$ phase diagram and red lines indicate the applied field strengths of 6.2 and 6.8 T, respectively.}
\label{Fig4}
\end{figure*}

After comparing $\Delta \theta_\text{MLD}(t)$ in the zigzag AFM ordered and QPM phases, we now turn to temperature dependent high fluence measurements at 6.2 and 6.8 T ($F=$\SI{80}{\micro\joule/\cm^2})(see Fig.~\ref{Fig4}(a)). Higher field measurements allow us to understand the nature of the long-lived plateauish state under higher pump laser excitation and to detect the QPM contribution in $\Delta \theta_\text{MLD}(t)$ more evidently. It is crucial to understand, whether there is a significant effect of laser induced heating on the transient magneto-optical response. If 
the heating effect is negligible, a temperature induced crossover from the AFM to the QPM upon increasing the bath temperature should be visible in the MLD transients. As the bath temperature is raised the sign of $\Delta \theta_\text{MLD}(t)$ changes from negative to positive. We interpret this behavior as being indicative of the thermal crossover from the zigzag AFM towards the QPM phase (cf. Fig.~\ref{Fig2}(a),(c)). Further, in the early dynamics, one can immediately recognize a mixture of two dynamical components, which motivated us to apply a multi-exponential fit with two distinct dynamical components related to the AFM and QPM state. 

\begin{equation}
\begin{split}
    \Delta\theta_{\text{MLD}}(t)&=A_\text{AFM}\,  (1-\mathrm{exp}(-t/\tau_\text{AFM}))\cdot\exp(-t/\tau_1)\\\
    &+A_\text{QPM}\,  (1-\mathrm{exp}(-t/\tau_\text{QPM}))\cdot\exp(-t/\tau_2)
\end{split}
\label{eq1}
\end{equation}

Here, $A_\text{AFM}$, $A_\text{QPM}$ correspond to the contribution of the AFM and QPM phase to the amplitude of the transient MLD; $\tau_\text{AFM}$, $\tau_\text{QPM}$ display the demagnetization time-constants for the two distinct phases and $\tau_1$,$\tau_2$ are added to capture the recovery dynamics on the 100s of ps time scales. An additional Gaussian response function capturing the laser profile was added, but does not influence the fit due to the different time scales. 
Interestingly, the time constants $\tau_\text{AFM}$ and $\tau_\text{QPM}$ of the two distinct states differ significantly, while being temperature-insensitive. We obtained $\sim5$ ps and $\sim10$ ps time constants for AFM and QPM demagnetization, respectively. This observation indicates that the system is optically driven into a non-thermal state. In addition, the extracted amplitudes of the fit (shown in Fig.~\ref{Fig4}(c)) show an interesting behavior. While the component related to the zigzag AFM dynamics vanishes approaching a temperature of $\approx 5$ K for an applied field strength of $6.2$ T, the QPM component is temperature independent. The vanishing of the AFM component at around 5 K is perfectly in line with static MLD measurements (c.f. Fig.~\ref{Fig4}(d)) indicating negligible laser heating effects even at the highest pump fluences. These findings indicate that the overall pump induced heating is negligible on macroscopic scales and that we observe the emergence of a photo-induced bi-partite phase, which is non-accessible by just tuning the bath temperature. These observations support the interpretation that the long-lived state at high fluences is not appearing due to a change in the temperature of the system but has its origin in the transient photo-induced change of the electronic and magnetic structure. Similar phase-coexistence in the vicinity of magnetic phase transitions have been observed previously. The mixture of two dynamical components far from equilibrium has been understood in terms of transiently quenched magnetic interactions by laser irradiation, which triggers the local formation and nucleation of magnetic domains of the other thermodynamic phase \cite{radu2010laser}. 

Having established the non-thermal nature of the fluence dependent $\Delta \theta_\text{MLD}(t)$ shown in Fig.~\ref{Fig2}(a), we now turn to discuss the possible underlying mechanisms and the origin of the long-lived state. The initial demagnetization time is on the order of 10s of ps and no faster component on sub-ps time scales has been observed in our measurements.
Previous magneto-optical studies on magnetically ordered systems reported similar time scales for an initial photo-induced demagnetization process \cite{afanasiev2016control,bergman2006identifying}, although the relevance of the involved microscopic degrees of freedom in different systems lead to a wide spreading of time scales in the literature \cite{koopmans2010explaining}. Recent 2PPES \cite{nevola2021timescales} directly revealed a sub-ps lifetime of doublons in the UHB after photo-excitation across the Mott gap, which is initially unexpected for such a large Mott-gapped insulator. In the same study some signatures of sub ps creation of Mott-Hubbard excitons have been reported, which has been interpreted as a reason of the fast decay of the doublons in the UHB in line with \cite{lenarvcivc2013ultrafast}. This fact underlines that the transient magnetic linear dichroism signal monitors the magnetic dynamics driven by the photo-excitations rather than the purely electronic response of the photo-excited charge carriers in the UHB. To reveal the underlying processes, which lead to the demagnetization dynamics displayed by the transient magnetic linear dichroism, the here presented comprehensive study under different experimental conditions, i.e., varying the applied pump fluence, bath temperature and external applied magnetic field, is crucial. 
At this point, the formation of holon-doublon pairs and Mott-Hubbard excitons is important, which have been observed in different Mott-insulating systems with an AFM ordered groundstate \cite{lovinger2020influence,afanasiev2019ultrafast,pastor2021non}. A recent study on the spin-orbit coupled Mott insulator Sr$_3$Ir$_2$O$_7$ \cite{pastor2021non} revealed that the photo-induced density of spin-less doublons is indeed crucial in understanding the dynamics after demagnetization. As mentioned above the photodoping of a Mott insulator creates holons and doublons in the LHB and UHB, respectively. By this charge excitation multiple spinless quasiparticles are created, such that magnetic moments are effectively removed from the lattice. When these charge excitations hop through the lattice, they continuously perturb and destroy the magnetic order \cite{li2020ultrafast}. This can be interpreted as creating spin defects, which naturally quench the zigzag order locally and hence decrease the zigzag antiferromagnetic order parameter. The strong charge-spin coupling has been theoretically proposed to display an efficient nonthermal demagnetization process \cite{balzer2015nonthermal}. In this regard, the fluence dependent MLD transients reveal the influence of the photo-excited holon-doublon density on the zigzag ordered state. At low excitation densities the changes in the orbital occupancies given by the present holon and doublons induce only minor spin defects and hence perturbations to the zigzag antiferromagnetic background. Any local frustration in the zigzag order given by misaligned pseudospins decreases the effective hopping between the sites. Hence, the surrounding intact zigzag order prevents the holons and doublons to move around, since this is energetically unfavorable in terms of the magnetic correlations. This is displayed by the experimental data, where no or short lived plateauish state can be observed at low excitation fluences, but the photo-induced changes in the transient MLD recover (c.f. Fig.~\ref{Fig2}(a)), i.e., the zigzag order is not perturbed beyond the initially created spin defects. Moreover, the amount of emitted magnons due to the excess energy during the recombination of holon doublon pairs does not suffice to fully spin disorder the system. Beyond the creation of holons and doublons it is theoretically predicted that the photodoping results in enhanced initial excitonic correlations, the behavior opposite to that caused by simple heating \cite{bittner2020photoenhanced}. It has been argued that long-lived excitations are generic for Mott-Hubbard insulators with large $U$ giving rise to the excitonic Mott-Hubbard insulator rather than a transient metallic state after photodoping \cite{eckstein2013photoinduced}, which has been commonly observed for narrow gap excitonic- and Mott-insulating materials \cite{okazaki2018photo,perfetti2008femtosecond,morrison2014photoinduced}. The Hubbard $U$ in $\alpha$-RuCl$_3$ has been reported to take a value of $U\approx4.35$ eV \cite{sinn2016electronic}. In $\alpha$-RuCl$_3$ the holons and doublons are found to form indeed an excitonic state in the Mott gap due to direct Coulomb interactions with a long-lived character, which is in accord with theory predicting in-gap states to be longer lived compared to short-lived excitonic states within the doublon-holon continuum \cite{bittner2020photoenhanced}. Different to the hopping of single holons and doublons the movement of holons and doublons paired up to a Mott-Hubbard exciton does not lead to a further disordering of the zigzag ordered groundstate, hence leaving zero net disorder behind, and they only contribute to the magnetization through magnon emission in the recombination process. At the time, the presence of Mott-Hubbard excitons can be regarded as a bottleneck for the reestablishment of long-range antiferromagnetic zigzag order, as in $\alpha$-RuCl$_3$ the holon-doublon bound together to an exciton are long-lived with lifetimes exceeding 100s of ps \cite{nevola2021timescales}. Thinking of an antiferromagnetically disordered zigzag groundstate by the presence of Mott-Hubbard excitons with long lifetime and breaking to single holons and doublons is in good agreement with the start of the recovery process at long times for low pump fluences. For increasing fluence, i.e., a higher excitation density the screening effect leads to a faster and larger amount of spin defects, the dynamics becomes different. As discussed above, the power law scaling of the initial demagnetization time $\tau_D$ with the fluence might indicate a breakdown of the Mott-Hubbard excitons, such that the number of free holons and doublons increases. Theory confirms that in the strong excitation regime a larger fraction of the excitons separates into unbound doublons and holons \cite{bittner2020photoenhanced}. For a larger amount of subsequent spin defects the zigzag order is already strongly perturbed, such that the holons and doublons then move easier through the lattice. In this regard they can separate further leaving a trace of disordered spins behind. In this way, the spin disordered state can be understood as a metastable state driven by the photo-induced holon-doublon density. In analogy, it has been reported for the magnetoresistive pyrochlore Tl$_2$Mn$_2$O$_7$, that the temporal persistence of the photoexcited carrier density can be strongly influenced by spin disorder \cite{prasankumar2005coupled}. For a certain amount of spin defects the long-range antiferromagnetic order cannot be reestablished by the magnetic exchange interactions. Hence, the fluence dependent demagnetization time fits to the picture of initially created spin defects causing a faster disordering of the zigzag order accompanied by a dramatic increase in the lifetime of the disordered state, which is in line with our observations. Monte Carlo simulations of a simple 3D spin-$1/2$ system showed that the amount of spin-flips, treated as the creation of double occupied sites, has significant influence on the recovery dynamics. For weak perturbations a quick recovery is seen, whereas the recovery dramatically slows down upon strong perturbations and a long-lived spin-disordered state appears \cite{pastor2021non}.


\begin{figure*}
\includegraphics[scale=0.6]{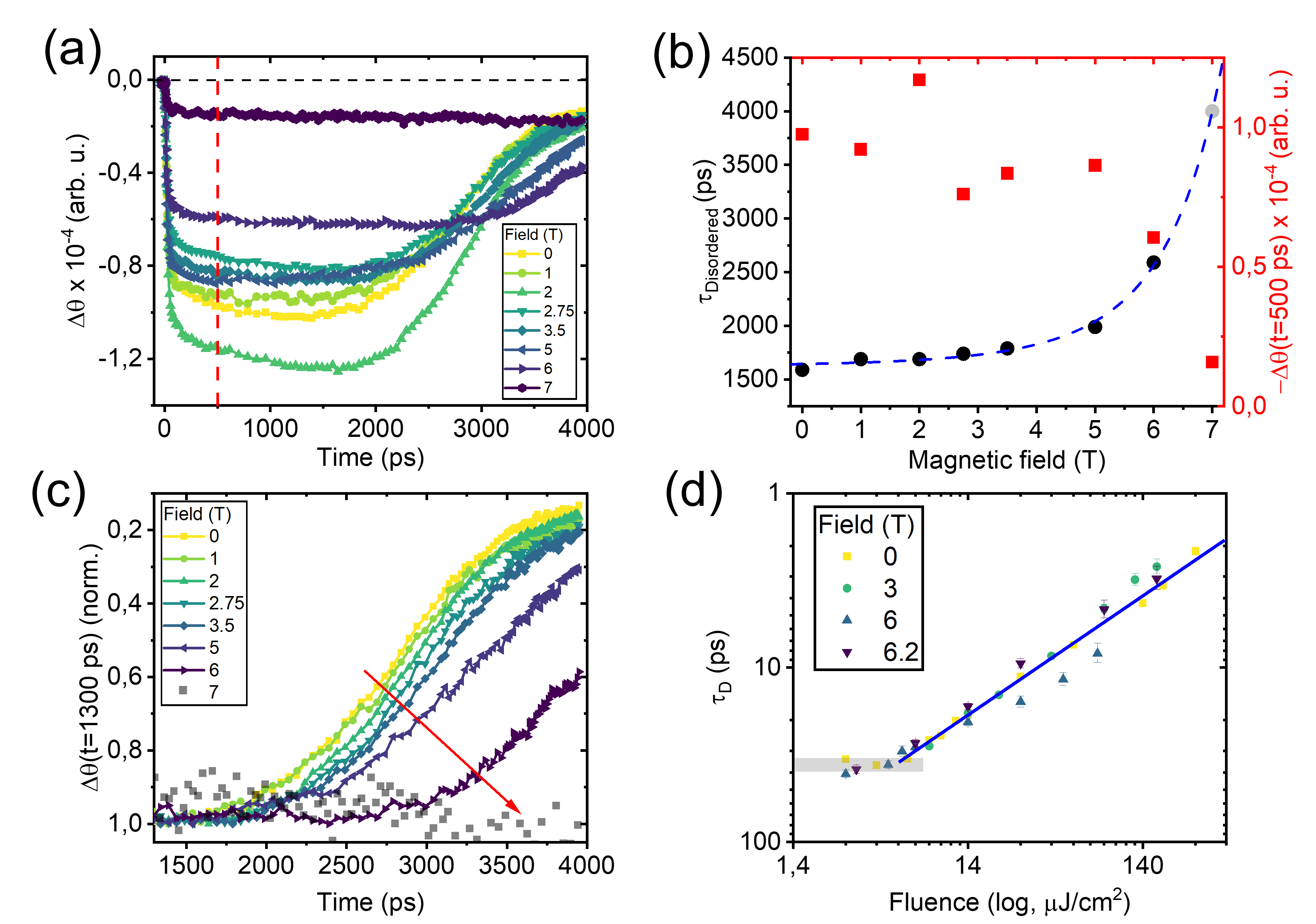}
\caption{{\bfseries{Field-dependent magneto-optical response in the zigzag AFM phase}}. (a) Field dependence of $\theta_\text{MLD}(t)$ at $3$ K for an applied pump fluences of \SI{15}{\micro\joule/\cm^2}. (b) Extracted changes in $\theta_\text{MLD}(t=500~\text{ps})$ (red squares), which follows the static magneto-optical response. The black circles indicate the lifetime of the plateau, which diverges towards approaching the equilibrium critical field strength above which $\alpha$-RuCl$_3$ enters the quantum disordered paramagnetic phase. (c) Normalized transients of $\Delta \theta(t=1300~\text{ps})$ shown in (a) for the long time delays. The red arrow indicates a slowing down of the recovery dynamics. (d) Fluence dependent demagnetization constants $\tau_D$ at different applied field strength on a log-log scale. The power law scaling discussed in the main text is indicated by the blue solid line.}
\label{Fig3}
\end{figure*}

Next, we turn to the field-dependence of $\Delta \theta_\text{MLD}(t)$ at a temperature of 3 K and an applied pump fluence of $F=$\SI{15}{\micro\joule/\cm^2}, which is shown in Fig.~\ref{Fig3}(a). One can clearly see, that in general the total change in $\Delta \theta_\text{MLD}(t)$ decreases with increasing applied magnetic field, which can be naturally assigned to the field-driven decrease of the order parameter. In small magnetic fields the dynamics is very similar, while for fields larger $2$ T the signal amplitude $\Delta \theta_\text{MLD}(t)$ changes and the plateau observed in the fluence dependence data in 0 T starts to become longer lived. Recent measurements found that at zero magnetic field the zigzag ground state is degenerate, i.e., there can be three differently oriented zigzag domains with the zigzag chain directions deviating by $120^\circ$ from each other. Applying a small in-plane magnetic field induces a change in the domain population that lifts the degeneracy and fits the observations of an initial small change in $|\Delta\theta(t)|$ between 0 and 1 T. In addition, under an in-plane magnetic field oriented perpendicular to one of the symmetry equivalent $\{1,1,0\}$ axis a field-driven metamagnetic transition has been identified at around 1.5 to 2 T \cite{wagnerstaticMLD}. Interestingly, the amplitude changes in  $\Delta \theta_\text{MLD}(t=500~\text{ps})$ displayed in Fig.~\ref{Fig3}(b) show an initial rise and sharp drop at around 2 T, which is related to the domain reorientation and clearly matches the reported hysteresis in the zigzag phase \cite{wagnerstaticMLD}, in line with expectations.
In addition, the lifetime of the metastable plateauish regime tends to diverge for increasing field strength approaching the field-driven phase transition boundary becoming longer lived than the accessible time delay, such that a lower bound of 4 ns at a field of 7 T has been assigned (see Fig.~\ref{Fig3}(b)). As the external applied field approaches the critical field value, magnetic correlations get suppressed and fluctuations in the order parameter occur in the vicinity of the phase transition \cite{dolgirev2020self}. In this picture holons and doublons then move on a many body fluctuating background with decreased magnetic correlations, which hinders the reestablishment of magnetic long-range order. Besides the longer lived spin-disordered phase also the recovery dynamics under an in-plane applied magnetic field slow down as shown in the normalized transient $\Delta \theta(t=1300 \text{ps})$ in Fig.~\ref{Fig3}(c). In contrast, the initial fluence dependent demagnetization dynamics show the power law scaling $\propto F^{-0.7}$ for all applied magnetic field strengths (see Fig.~\ref{Fig3}(d). This provides another proof for our initial discussion about the purely electronic mechanism behind the demagnetization. The holon-doublon density is indeed crucial in understanding the photo-induced quench of the zigzag antiferromagnetic state. Up to a threshold fluence $\tau_D$ is constant, which points towards the initial formation of Mott-Hubbard excitons by the photo-excited holons and doublons referring to \cite{nevola2021timescales}. Above this threshold fluence the amount of excess free holons and doublons increases and causes a screening effect, such that the formed excitons break up into free holons and doublons. This corresponds to a larger amount of spin-defects disturbing the AFM background, which leads to the emergence of the long-lived plateauish regime in $\Delta\theta_\text{MLD}(t)$.

From the experimental data, we can conclude that the field dependent transient MLD responses for the zigzag AFM and the QPM phase are intrinsically different. Considering the results for the field-aligned QPM phase (cf. Fig.~\ref{Fig2}(b)) we can make the following statements. Although the excitation mechanism above and below $T_N$ is the same, the difference in the demagnetization dynamics indicates clearly that the presence of antiferromagnetic correlations influences the dynamics of photo-excited holons and doublons significantly and is crucially in understanding the formation of a long-lived spin-disordered state. From inhomogeneous nonequilibrium dynamical mean-field theory, it is predicted that antiferromagnetic correlations indeed influence charge excitations \cite{eckstein2014ultrafast}. Based on our experimental data we conclude that the presence of AFM long range interactions restricts the movement of holons and doublons for sufficiently low excitation densities. Further, a certain amount of holons and doublons needs to be created to break up the already formed excitons and to decrease the AFM correlations via the creation of a high density of local spin defects. In addition, the magnetic field also reduces AFM correlations and leads to a slowing down of the recovery dynamics in the zigzag AFM phase as shown in Fig.~\ref{Fig3}(c). In contrast, the recovery of $\Delta \theta_\text{MLD}(t)$ at $9$ K is similar for all applied field strengths and does not show a field dependence. This is another indicator that the recovery dynamics are influenced by the interaction between the holons and doublons and the underlying long range magnetic order.

\begin{figure*}
\includegraphics[scale=0.5]{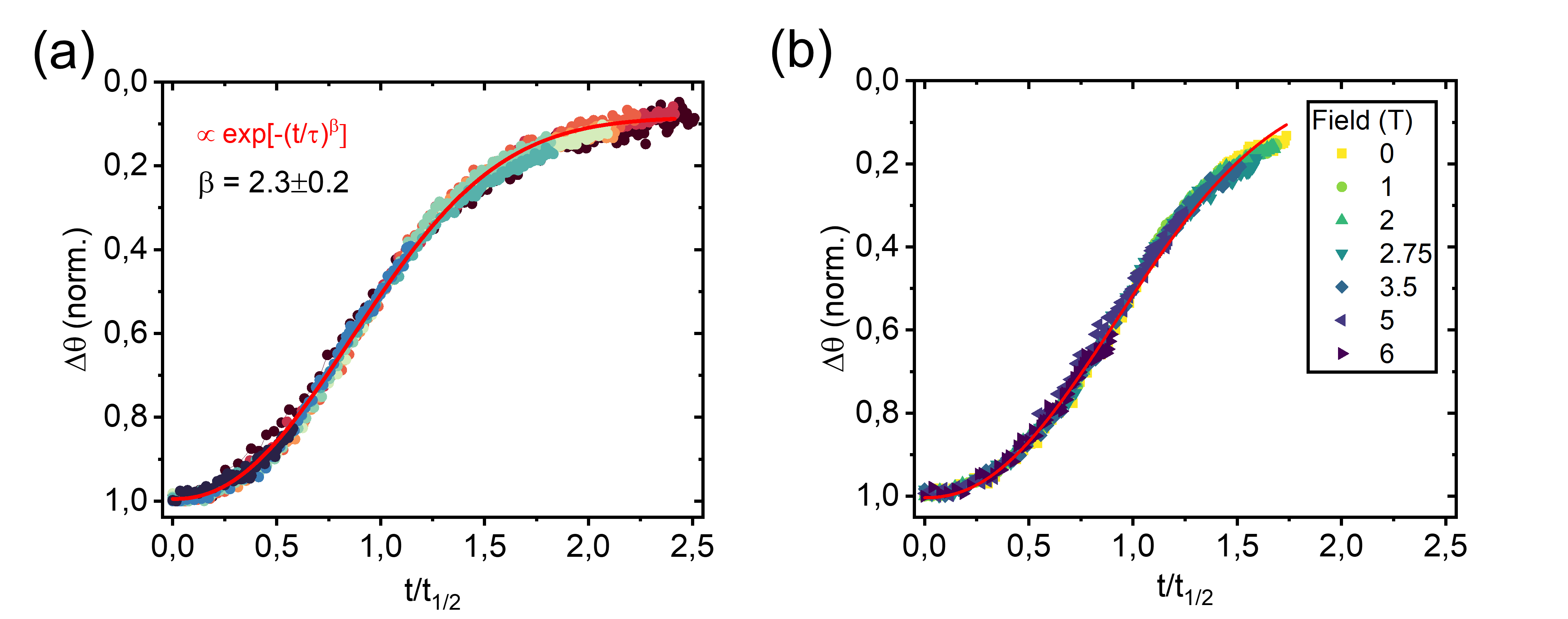}
\caption{{\bfseries{Universal recovery dynamics in the zigzag AFM phase}}. (a) Rescaled fluence dependence of the long-time recovery $\theta_\text{MLD}(t)$ in zero magnetic field at $3$ K for applied pump fluences varying between $3$ to \SI{250}{\micro\joule/\cm^2}. (b) Rescaled field dependence of the long-time recovery $\theta_\text{MLD}(t)$ at $3$ K for applied pump fluences of \SI{15}{\micro\joule/\cm^2}. The data in (a) and (b) is normalized to its value at which the recovery process starts and the time axis is rescaled to $t_{1/2}$ as discussed in the main text. The red solid line indicates a compressed exponential fit associated with domain propagation dynamics.}
\label{Fig5}
\end{figure*}

Up to now we have mainly discussed the demagnetization and the induced metastable phase. We now turn to the recovery, we discuss the recovery dynamics of the transient magnetic linear dichroism, which displays how the photo-induced magnetically disordered state recovers back to its pre-excitation equilibrium state. Differently from the observed non-linear fluence dependence of the initial demagnetization dynamics, the fluence dependence of the recovery dynamics indicates a universal scaling behavior. Although the starting point of the recovery process is non-linear in the applied fluence and show a field dependent behavior due to the reduced antiferromagnetic background (Fig.~\ref{Fig4}(c)), in the following, we argue that the \textit{mechanism} of recovery is similar for all measured conditions. 
In Fig.~\ref{Fig5}(a) we show the rescaled long-time recovery of $\Delta \theta_\text{MLD}(t)$ at 2.5 K and zero applied field. The MLD data has been normalized to its amplitude value at which the recovery process starts and the time axis has been rescaled by $t_{1/2}$, the time at which $\Delta \theta_\text{MLD}(t)$ has already recovered by 50 $\%$. As can be seen clearly, the rescaled transients fall on top of each other pointing towards an underlying universal decay process. The recovery dynamics can be best fit using a compressed exponential decay function $\propto \exp{(-(t/\tau)^\beta)}$, with an extracted fluence-independent exponent $\beta=(2.3\pm0.2)$. The recovery time constant $\sim1$ ns of the unscaled data transients matches the time scale of slow diffusion driven processes, which are often observed in pump-probe experiments, when the heat in the excited probe area is transferred along a temperature gradient to unpumped, colder regions. Although we cannot exclude the presence of diffusion driven recovery, the fluence-independent compressed exponential behavior points to a minor heating effect, which should otherwise influence the dynamics significantly. A compressed exponential behavior has been observed in the dynamics of antiferromagnetic domain wall dynamics \cite{shpyrko2007direct}, for a variety of soft matter systems undergoing ‘jamming’ transitions, disordered glassy states and has also been interpreted in terms of ballistic motion of elastic deformation in response to heterogeneous local stress \cite{cipelletti2003universal}. Further, the phenomenon of compressed exponential relaxation has been discussed in terms of magnetization reversal in magnetic thin films \cite{xi2008slow}. Here, the Fatuzzo-Labrune model describes magnetization reversal in a simple and intuitive way in terms of an initial and probabilistic nucleation process, which is followed by either further nucleation governed magnetization reversal or by a domain wall propagation driven process. Although, the Fatuzzo-Labrune model makes an oversimplification of the present magnetic domains treating them as circularly shaped, it can give useful qualitative information. The time evolution of the dynamics is generally captured by $\exp(-(t/\tau)^\beta)$, while the value of $\beta$ gives information about the underlying recovery processes \cite{labrune1989time,xi2008slow} and allows to distinguish between domain wall propagation and nucleation driven relaxation. It was established that the shape of the relaxation curves depends on the relative importance of domain wall motion and nucleation processes. The relaxation curves show an S-shape (compressed exponential) for relaxation dynamics dominated by domain wall propagation ($\beta>1$), while for nucleation dominated reversal the relaxation curve takes a (stretched) exponential shape ($\beta\leq1$) \cite{romanens2005magnetic,xi2008slow,labrune1989time}.

Therefore, examination of the shape of the scaled recovery transient curves suggests that the reversal is mainly governed by domain wall propagation (see Fig.~\ref{Fig5}). Normally, magnetization reversal is a slow process on time scales extending up to several seconds or even minutes \cite{adjanoh2011compressed,pommier1990magnetization}. However, magnetization reversal in these studies is usually triggered by an external applied magnetic field under purely equilibrium conditions. Here, the situation is drastically different since we report on the dynamics of the magnetic sector under nonequilibrium conditions. This might pave the way to understand the clearly different time scales we observe here. 
The pump laser excitation quenches the zigzag antiferromagnetic order locally within the area of the laser spot. A metastable state is created, which is, as discussed above, mainly described by the presence of holons and doublons interpreted as local spin defects in the zigzag phase. The compressed exponential clearly indicates a relaxation process driven by domain wall propagation within the here proposed model. The picture of domain wall propagation also holds under applied magnetic fields as can be deduced from Fig.~\ref{Fig5}(b). The recovery dynamics in field can also be rescaled and follow a compressed exponential recovery with an exponent of $\beta=(2.21\pm0.04$) in line with the fluence dependence recovery in 0 applied magnetic field. Hence, it seems that after a certain time the zigzag AFM order builds up locally again, i.e. a first zigzag domain is formed within a spin-disordered state. This domain then expands in time, i.e. its domain wall propagates throughout the probed area in contrast to a multiple domain nucleation scenario. Which process helps to form this initial starting domain is not evident. The presence of local defects and domain wall pinning are considerable \cite{pommier1990magnetization}, but experimental probes, which visualize the domain structure directly \cite{cheong2020seeing} are necessary to address this question in more detail. It is interesting, that the temporal behavior governed by a compressed exponential decay matches the observations for magnetic reordering processes in thermodynamic equilibrium pretty well. This interpretation supports our argument about the strong effect of magnetic background on the fast recovery in the LFR and the long-lived plateau in the IFR. Here the domain wall propagation is indicative of locally strong antiferromagnetic order which assists holons and doublons to relax and becomes dominant after initial random nucleation in the beginning of the recovery process.

\section{Summary and conclusions\label{Conclusions}}

In this report, we presented the time-resolved dynamics of the spin liquid
candidate $\alpha$-RuCl$_3$ throughout the whole low-temperature zigzag antiferromagnetically ordered phase upon photo-excitation above the Mott-Hubbard gap. The photo-excitation creates holons and doublons in the lower and upper Hubbard band, respectively. Due to strong charge-spin coupling in the spin-orbit assisted Mott insulator, the effect of charge excitations on the antiferromagnetically ordered state can be efficiently probed via transient magnetic linear dichroism. Our measurements reveal that the density of these spinless quasiparticles opens a nonequilibrium way to destabilize the zigzag antiferromagnetic order and can be treated as an additional dimension to the $\alpha$-RuCl$_3$ phase diagram to create a non-thermal photo-induced long-lived and spin-disordered state. The presence of holons and doublons and their effect on the antiferromagnetically ordered background can be understood within a picture of spin defects that move around and disturb the zigzag order. Further, our measurements indirectly point towards the presence of holon-doublon pairs forming Mott-Hubbard excitons with 30 ps lifetime at low excitation densities. The results show the demagnetization is a consequence of the relaxation of excited holons and doublons and it occurs through competing mechanisms, either magnon emission or hopping of unbound carriers. The dominating demagnetization process seems to be dependent on the excitation density. Whereas, for small excitation densities the exciton recombination via magnon emission is dominant, the hopping of unbound holons and doublons effectitvely perturbs the zigzag order upon high excitation densities. Further, the magnetic field dependent transient magnetic linear dichroism shows pseudocritical behavior, especially a longer lifetime of the photoinduced metastable state reminiscent of the field driven decrease of zigzag antiferromagnetic correlations. 
Besides the creation and nature of the long-lived spin-disordered state, our study also indicates that the recovery dynamics can be understood in terms of a magnetic reordering process rather than simple heat diffusion driving thermalization of the magnetic, lattice and charge degrees of freedom. The recovery dynamics follow a universal compressed exponential form with a compressing exponent $\beta=(2.3\pm0.2)$ indicative for a domain wall propagation dominated reordering process.\\
Our study revealed that photoexcitation of charge carriers can be used to destabilize magnetic order in systems with strongly intertwined degrees of freedom like $\alpha$-RuCl$_3$. It is interesting to study other spin liquid candidates (e.g. Na$_2$IrO$_3$ \cite{chaloupka2016magnetic}) and other magnetic Mott insulators to clarify that the presence of charge excitations does play a critical role in stabilizing hidden phases, helps to quench antiferromagnetic order and to access metastable non-thermal states. In this regard, there remain still some important open questions that need to be addressed in future experiments. First, a broad spectral probing of the magneto-optical response would give a deeper understanding of the interactions between the charge, lattice and magnetic sectors. Further, it would be interesting to study the electronic and lattice dynamics at low temperatures and long delays more directly by complementary time-resolved techniques \cite{hedayat2021investigation}, like time-resolved ARPES \cite{hedayat2018surface}, XRD \cite{gerber2017femtosecond}, Raman \cite{versteeg2018tunable}, etc. These studies will reveal the intrinsic time scales and interactions of holons and doublons and other quasiparticles and could answer whether a transient lattice deformation can quench magnetic exchange interactions, while simultaneously enhancing the anisotropic Kitaev interactions to drive $\alpha$-RuCl$_3$ into the Kitaev spin liquid state. Moreover, it would be important to understand theoretically the qualitative correspondence between the here observed compressed exponential decay under nonequilibrium conditions and magnetization reversal in ferromagnets in equilibrium. 

\section*{Conflicts of interest}
There are no conflicts to declare.

\section*{Acknowledgments}
We are indebted to Fulvio Parmigiani, Daniel I. Khomskii, Sebastian Diehl and Alessio Chiocchetta for fruitful discussions. The authors acknowledge financial support funded by the Deutsche Forschungsgemeinschaft (DFG) through project No. 277146847-CRC1238, Control and Dynamics of Quantum Materials (subproject No. B05). AL and VT acknowledge support by the Deutsche Forschungsgemeinschaft (DFG) through the Transregional Research Collaboration TRR 80$:$ From Electronic Correlations to Functionality (Augsburg, Munich, and Stuttgart). VT acknowledges the support via the project ANCD 20.80009.5007.19 (Moldova).\\


\def\bibsection{\section*{~\refname}} 
\bibliography{bibliography}

\providecommand{\noopsort}[1]{}\providecommand{\singleletter}[1]{#1}
\begin{thebibliography}{73}%
\makeatletter
\providecommand \@ifxundefined [1]{%
 \@ifx{#1\undefined}
}%
\providecommand \@ifnum [1]{%
 \ifnum #1\expandafter \@firstoftwo
 \else \expandafter \@secondoftwo
 \fi
}%
\providecommand \@ifx [1]{%
 \ifx #1\expandafter \@firstoftwo
 \else \expandafter \@secondoftwo
 \fi
}%
\providecommand \natexlab [1]{#1}%
\providecommand \enquote  [1]{``#1''}%
\providecommand \bibnamefont  [1]{#1}%
\providecommand \bibfnamefont [1]{#1}%
\providecommand \citenamefont [1]{#1}%
\providecommand \href@noop [0]{\@secondoftwo}%
\providecommand \href [0]{\begingroup \@sanitize@url \@href}%
\providecommand \@href[1]{\@@startlink{#1}\@@href}%
\providecommand \@@href[1]{\endgroup#1\@@endlink}%
\providecommand \@sanitize@url [0]{\catcode `\\12\catcode `\$12\catcode
  `\&12\catcode `\#12\catcode `\^12\catcode `\_12\catcode `\%12\relax}%
\providecommand \@@startlink[1]{}%
\providecommand \@@endlink[0]{}%
\providecommand \url  [0]{\begingroup\@sanitize@url \@url }%
\providecommand \@url [1]{\endgroup\@href {#1}{\urlprefix }}%
\providecommand \urlprefix  [0]{URL }%
\providecommand \Eprint [0]{\href }%
\providecommand \doibase [0]{https://doi.org/}%
\providecommand \selectlanguage [0]{\@gobble}%
\providecommand \bibinfo  [0]{\@secondoftwo}%
\providecommand \bibfield  [0]{\@secondoftwo}%
\providecommand \translation [1]{[#1]}%
\providecommand \BibitemOpen [0]{}%
\providecommand \bibitemStop [0]{}%
\providecommand \bibitemNoStop [0]{.\EOS\space}%
\providecommand \EOS [0]{\spacefactor3000\relax}%
\providecommand \BibitemShut  [1]{\csname bibitem#1\endcsname}%
\let\auto@bib@innerbib\@empty
\bibitem [{\citenamefont {de~la Torre}\ \emph {et~al.}(2021)\citenamefont
  {de~la Torre}, \citenamefont {Kennes}, \citenamefont {Claassen},
  \citenamefont {Gerber}, \citenamefont {McIver},\ and\ \citenamefont
  {Sentef}}]{de2021nonthermal}%
  \BibitemOpen
  \bibfield  {author} {\bibinfo {author} {\bibfnamefont {A.}~\bibnamefont
  {de~la Torre}}, \bibinfo {author} {\bibfnamefont {D.~M.}\ \bibnamefont
  {Kennes}}, \bibinfo {author} {\bibfnamefont {M.}~\bibnamefont {Claassen}},
  \bibinfo {author} {\bibfnamefont {S.}~\bibnamefont {Gerber}}, \bibinfo
  {author} {\bibfnamefont {J.~W.}\ \bibnamefont {McIver}},\ and\ \bibinfo
  {author} {\bibfnamefont {M.~A.}\ \bibnamefont {Sentef}},\ }\bibfield  {title}
  {\bibinfo {title} {Nonthermal pathways to ultrafast control in quantum
  materials},\ }\href@noop {} {\bibfield  {journal} {\bibinfo  {journal} {arXiv
  preprint arXiv:2103.14888}\ } (\bibinfo {year} {2021})}\BibitemShut {NoStop}%
\bibitem [{\citenamefont {Li}\ \emph {et~al.}(2018)\citenamefont {Li},
  \citenamefont {Strand}, \citenamefont {Werner},\ and\ \citenamefont
  {Eckstein}}]{li2018theory}%
  \BibitemOpen
  \bibfield  {author} {\bibinfo {author} {\bibfnamefont {J.}~\bibnamefont
  {Li}}, \bibinfo {author} {\bibfnamefont {H.~U.}\ \bibnamefont {Strand}},
  \bibinfo {author} {\bibfnamefont {P.}~\bibnamefont {Werner}},\ and\ \bibinfo
  {author} {\bibfnamefont {M.}~\bibnamefont {Eckstein}},\ }\bibfield  {title}
  {\bibinfo {title} {Theory of photoinduced ultrafast switching to a
  spin-orbital ordered hidden phase},\ }\href@noop {} {\bibfield  {journal}
  {\bibinfo  {journal} {Nature communications}\ }\textbf {\bibinfo {volume}
  {9}},\ \bibinfo {pages} {1} (\bibinfo {year} {2018})}\BibitemShut {NoStop}%
\bibitem [{\citenamefont {Kalashnikova}\ \emph {et~al.}(2015)\citenamefont
  {Kalashnikova}, \citenamefont {Kimel},\ and\ \citenamefont
  {Pisarev}}]{kalashnikova2015ultrafast}%
  \BibitemOpen
  \bibfield  {author} {\bibinfo {author} {\bibfnamefont {A.~M.}\ \bibnamefont
  {Kalashnikova}}, \bibinfo {author} {\bibfnamefont {A.~V.}\ \bibnamefont
  {Kimel}},\ and\ \bibinfo {author} {\bibfnamefont {R.~V.}\ \bibnamefont
  {Pisarev}},\ }\bibfield  {title} {\bibinfo {title} {Ultrafast
  opto-magnetism},\ }\href@noop {} {\bibfield  {journal} {\bibinfo  {journal}
  {Physics-Uspekhi}\ }\textbf {\bibinfo {volume} {58}},\ \bibinfo {pages} {969}
  (\bibinfo {year} {2015})}\BibitemShut {NoStop}%
\bibitem [{\citenamefont {Kirilyuk}\ \emph {et~al.}(2010)\citenamefont
  {Kirilyuk}, \citenamefont {Kimel},\ and\ \citenamefont
  {Rasing}}]{kirilyuk2010ultrafast}%
  \BibitemOpen
  \bibfield  {author} {\bibinfo {author} {\bibfnamefont {A.}~\bibnamefont
  {Kirilyuk}}, \bibinfo {author} {\bibfnamefont {A.~V.}\ \bibnamefont
  {Kimel}},\ and\ \bibinfo {author} {\bibfnamefont {T.}~\bibnamefont
  {Rasing}},\ }\bibfield  {title} {\bibinfo {title} {Ultrafast optical
  manipulation of magnetic order},\ }\href@noop {} {\bibfield  {journal}
  {\bibinfo  {journal} {Reviews of Modern Physics}\ }\textbf {\bibinfo {volume}
  {82}},\ \bibinfo {pages} {2731} (\bibinfo {year} {2010})}\BibitemShut
  {NoStop}%
\bibitem [{\citenamefont {Jackeli}\ and\ \citenamefont
  {Khaliullin}(2009)}]{jackeli2009mott}%
  \BibitemOpen
  \bibfield  {author} {\bibinfo {author} {\bibfnamefont {G.}~\bibnamefont
  {Jackeli}}\ and\ \bibinfo {author} {\bibfnamefont {G.}~\bibnamefont
  {Khaliullin}},\ }\bibfield  {title} {\bibinfo {title} {Mott insulators in the
  strong spin-orbit coupling limit: from {H}eisenberg to a quantum compass and
  {K}itaev models},\ }\href@noop {} {\bibfield  {journal} {\bibinfo  {journal}
  {Phys. Rev. Lett.}\ }\textbf {\bibinfo {volume} {102}},\ \bibinfo {pages}
  {017205} (\bibinfo {year} {2009})}\BibitemShut {NoStop}%
\bibitem [{\citenamefont {Savary}\ and\ \citenamefont
  {Balents}(2016)}]{savary2016quantum}%
  \BibitemOpen
  \bibfield  {author} {\bibinfo {author} {\bibfnamefont {L.}~\bibnamefont
  {Savary}}\ and\ \bibinfo {author} {\bibfnamefont {L.}~\bibnamefont
  {Balents}},\ }\bibfield  {title} {\bibinfo {title} {Quantum spin liquids: a
  review},\ }\href@noop {} {\bibfield  {journal} {\bibinfo  {journal} {Reports
  on Progress in Physics}\ }\textbf {\bibinfo {volume} {80}},\ \bibinfo {pages}
  {016502} (\bibinfo {year} {2016})}\BibitemShut {NoStop}%
\bibitem [{\citenamefont {{K}itaev}(2006)}]{Kitaev20062}%
  \BibitemOpen
  \bibfield  {author} {\bibinfo {author} {\bibfnamefont {A.}~\bibnamefont
  {{K}itaev}},\ }\bibfield  {title} {\bibinfo {title} {{A}nyons in an exactly
  solved model and beyond},\ }\href
  {https://doi.org/https://doi.org/10.1016/j.aop.2005.10.005} {\bibfield
  {journal} {\bibinfo  {journal} {Annals of Physics}\ }\textbf {\bibinfo
  {volume} {321}},\ \bibinfo {pages} {2} (\bibinfo {year} {2006})},\ \bibinfo
  {note} {january Special Issue}\BibitemShut {NoStop}%
\bibitem [{\citenamefont {Glamazda}\ \emph {et~al.}(2017)\citenamefont
  {Glamazda}, \citenamefont {Lemmens}, \citenamefont {Do}, \citenamefont
  {Kwon},\ and\ \citenamefont {Choi}}]{glamazda2017relation}%
  \BibitemOpen
  \bibfield  {author} {\bibinfo {author} {\bibfnamefont {A.}~\bibnamefont
  {Glamazda}}, \bibinfo {author} {\bibfnamefont {P.}~\bibnamefont {Lemmens}},
  \bibinfo {author} {\bibfnamefont {S.-H.}\ \bibnamefont {Do}}, \bibinfo
  {author} {\bibfnamefont {Y.}~\bibnamefont {Kwon}},\ and\ \bibinfo {author}
  {\bibfnamefont {K.-Y.}\ \bibnamefont {Choi}},\ }\bibfield  {title} {\bibinfo
  {title} {Relation between {K}itaev magnetism and structure in
  $\alpha$-{R}u{C}l$_3$},\ }\href@noop {} {\bibfield  {journal} {\bibinfo
  {journal} {Physical Review B}\ }\textbf {\bibinfo {volume} {95}},\ \bibinfo
  {pages} {174429} (\bibinfo {year} {2017})}\BibitemShut {NoStop}%
\bibitem [{\citenamefont {Loidl}\ \emph {et~al.}(2021)\citenamefont {Loidl},
  \citenamefont {Lunkenheimer},\ and\ \citenamefont
  {Tsurkan}}]{loidl2021proximate}%
  \BibitemOpen
  \bibfield  {author} {\bibinfo {author} {\bibfnamefont {A.}~\bibnamefont
  {Loidl}}, \bibinfo {author} {\bibfnamefont {P.}~\bibnamefont
  {Lunkenheimer}},\ and\ \bibinfo {author} {\bibfnamefont {V.}~\bibnamefont
  {Tsurkan}},\ }\bibfield  {title} {\bibinfo {title} {On the proximate {K}itaev
  quantum-spin liquid $\alpha$-{R}u{C}l$_3$: thermodynamics, excitations and
  continua},\ }\href@noop {} {\bibfield  {journal} {\bibinfo  {journal}
  {Journal of Physics: Condensed Matter}\ } (\bibinfo {year}
  {2021})}\BibitemShut {NoStop}%
\bibitem [{\citenamefont {Trebst}(2017)}]{trebst2017kitaev}%
  \BibitemOpen
  \bibfield  {author} {\bibinfo {author} {\bibfnamefont {S.}~\bibnamefont
  {Trebst}},\ }\href@noop {} {\bibinfo {title} {Kitaev materials}} (\bibinfo
  {year} {2017}),\ \Eprint {https://arxiv.org/abs/1701.07056} {arXiv:1701.07056
  [cond-mat.str-el]} \BibitemShut {NoStop}%
\bibitem [{\citenamefont {Kasahara}\ \emph {et~al.}(2018)\citenamefont
  {Kasahara}, \citenamefont {Ohnishi}, \citenamefont {Mizukami}, \citenamefont
  {Tanaka}, \citenamefont {Ma}, \citenamefont {Sugii}, \citenamefont {Kurita},
  \citenamefont {Tanaka}, \citenamefont {Nasu}, \citenamefont {Motome} \emph
  {et~al.}}]{kasahara2018majorana}%
  \BibitemOpen
  \bibfield  {author} {\bibinfo {author} {\bibfnamefont {Y.}~\bibnamefont
  {Kasahara}}, \bibinfo {author} {\bibfnamefont {T.}~\bibnamefont {Ohnishi}},
  \bibinfo {author} {\bibfnamefont {Y.}~\bibnamefont {Mizukami}}, \bibinfo
  {author} {\bibfnamefont {O.}~\bibnamefont {Tanaka}}, \bibinfo {author}
  {\bibfnamefont {S.}~\bibnamefont {Ma}}, \bibinfo {author} {\bibfnamefont
  {K.}~\bibnamefont {Sugii}}, \bibinfo {author} {\bibfnamefont
  {N.}~\bibnamefont {Kurita}}, \bibinfo {author} {\bibfnamefont
  {H.}~\bibnamefont {Tanaka}}, \bibinfo {author} {\bibfnamefont
  {J.}~\bibnamefont {Nasu}}, \bibinfo {author} {\bibfnamefont {Y.}~\bibnamefont
  {Motome}}, \emph {et~al.},\ }\bibfield  {title} {\bibinfo {title} {Majorana
  quantization and half-integer thermal quantum {H}all effect in a {K}itaev
  spin liquid},\ }\href@noop {} {\bibfield  {journal} {\bibinfo  {journal}
  {Nature}\ }\textbf {\bibinfo {volume} {559}},\ \bibinfo {pages} {227}
  (\bibinfo {year} {2018})}\BibitemShut {NoStop}%
\bibitem [{\citenamefont {Wen}\ \emph {et~al.}(2019)\citenamefont {Wen},
  \citenamefont {Yu}, \citenamefont {Li}, \citenamefont {Yu},\ and\
  \citenamefont {Li}}]{wen2019experimental}%
  \BibitemOpen
  \bibfield  {author} {\bibinfo {author} {\bibfnamefont {J.}~\bibnamefont
  {Wen}}, \bibinfo {author} {\bibfnamefont {S.-L.}\ \bibnamefont {Yu}},
  \bibinfo {author} {\bibfnamefont {S.}~\bibnamefont {Li}}, \bibinfo {author}
  {\bibfnamefont {W.}~\bibnamefont {Yu}},\ and\ \bibinfo {author}
  {\bibfnamefont {J.-X.}\ \bibnamefont {Li}},\ }\bibfield  {title} {\bibinfo
  {title} {Experimental identification of quantum spin liquids},\ }\href@noop
  {} {\bibfield  {journal} {\bibinfo  {journal} {npj Quantum Materials}\
  }\textbf {\bibinfo {volume} {4}},\ \bibinfo {pages} {1} (\bibinfo {year}
  {2019})}\BibitemShut {NoStop}%
\bibitem [{\citenamefont {Cao}\ \emph {et~al.}(2016)\citenamefont {Cao},
  \citenamefont {Banerjee}, \citenamefont {Yan}, \citenamefont {Bridges},
  \citenamefont {Lumsden}, \citenamefont {Mandrus}, \citenamefont {Tennant},
  \citenamefont {Chakoumakos},\ and\ \citenamefont {Nagler}}]{cao2016low}%
  \BibitemOpen
  \bibfield  {author} {\bibinfo {author} {\bibfnamefont {H.~B.}\ \bibnamefont
  {Cao}}, \bibinfo {author} {\bibfnamefont {A.}~\bibnamefont {Banerjee}},
  \bibinfo {author} {\bibfnamefont {J.-Q.}\ \bibnamefont {Yan}}, \bibinfo
  {author} {\bibfnamefont {C.}~\bibnamefont {Bridges}}, \bibinfo {author}
  {\bibfnamefont {M.}~\bibnamefont {Lumsden}}, \bibinfo {author} {\bibfnamefont
  {D.}~\bibnamefont {Mandrus}}, \bibinfo {author} {\bibfnamefont
  {D.}~\bibnamefont {Tennant}}, \bibinfo {author} {\bibfnamefont
  {B.}~\bibnamefont {Chakoumakos}},\ and\ \bibinfo {author} {\bibfnamefont
  {S.}~\bibnamefont {Nagler}},\ }\bibfield  {title} {\bibinfo {title}
  {Low-temperature crystal and magnetic structure of $\alpha$-{R}u{C}l$_3$},\
  }\href@noop {} {\bibfield  {journal} {\bibinfo  {journal} {Physical Review
  B}\ }\textbf {\bibinfo {volume} {93}},\ \bibinfo {pages} {134423} (\bibinfo
  {year} {2016})}\BibitemShut {NoStop}%
\bibitem [{\citenamefont {Wagner}\ \emph {et~al.}(2022)\citenamefont {Wagner},
  \citenamefont {Sahasrabudhe}, \citenamefont {Versteeg}, \citenamefont
  {Wysocki}, \citenamefont {Wang}, \citenamefont {Tsurkan}, \citenamefont
  {Loidl}, \citenamefont {Khomskii}, \citenamefont {Hedayat},\ and\
  \citenamefont {van Loosdrecht}}]{wagnerstaticMLD}%
  \BibitemOpen
  \bibfield  {author} {\bibinfo {author} {\bibfnamefont {J.}~\bibnamefont
  {Wagner}}, \bibinfo {author} {\bibfnamefont {A.}~\bibnamefont
  {Sahasrabudhe}}, \bibinfo {author} {\bibfnamefont {R.}~\bibnamefont
  {Versteeg}}, \bibinfo {author} {\bibfnamefont {L.}~\bibnamefont {Wysocki}},
  \bibinfo {author} {\bibfnamefont {Z.}~\bibnamefont {Wang}}, \bibinfo {author}
  {\bibfnamefont {V.}~\bibnamefont {Tsurkan}}, \bibinfo {author} {\bibfnamefont
  {A.}~\bibnamefont {Loidl}}, \bibinfo {author} {\bibfnamefont
  {D.}~\bibnamefont {Khomskii}}, \bibinfo {author} {\bibfnamefont
  {H.}~\bibnamefont {Hedayat}},\ and\ \bibinfo {author} {\bibfnamefont
  {P.}~\bibnamefont {van Loosdrecht}},\ }\href@noop {} {\bibinfo {title}
  {Magneto-optical study of metamagnetic transitions in the antiferromagnetic
  phase of $\alpha$-{R}u{C}l$_3$}} (\bibinfo {year} {2022}),\ \Eprint
  {https://arxiv.org/abs/2201.02842} {arXiv:2201.02842 [cond-mat.str-el]}
  \BibitemShut {NoStop}%
\bibitem [{\citenamefont {Fletcher}\ \emph {et~al.}(1967)\citenamefont
  {Fletcher}, \citenamefont {Gardner}, \citenamefont {Fox},\ and\ \citenamefont
  {Topping}}]{fletcher1967x}%
  \BibitemOpen
  \bibfield  {author} {\bibinfo {author} {\bibfnamefont {J.}~\bibnamefont
  {Fletcher}}, \bibinfo {author} {\bibfnamefont {W.}~\bibnamefont {Gardner}},
  \bibinfo {author} {\bibfnamefont {A.}~\bibnamefont {Fox}},\ and\ \bibinfo
  {author} {\bibfnamefont {G.}~\bibnamefont {Topping}},\ }\bibfield  {title}
  {\bibinfo {title} {X-ray, infrared, and magnetic studies of $\alpha$-and
  $\beta$-ruthenium trichloride},\ }\href@noop {} {\bibfield  {journal}
  {\bibinfo  {journal} {Journal of the Chemical Society A: Inorganic, Physical,
  Theoretical}\ ,\ \bibinfo {pages} {1038}} (\bibinfo {year}
  {1967})}\BibitemShut {NoStop}%
\bibitem [{\citenamefont {Sears}\ \emph {et~al.}(2020)\citenamefont {Sears},
  \citenamefont {Chern}, \citenamefont {Kim}, \citenamefont {Bereciartua},
  \citenamefont {Francoual}, \citenamefont {Kim},\ and\ \citenamefont
  {Kim}}]{sears2020ferromagnetic}%
  \BibitemOpen
  \bibfield  {author} {\bibinfo {author} {\bibfnamefont {J.~A.}\ \bibnamefont
  {Sears}}, \bibinfo {author} {\bibfnamefont {L.~E.}\ \bibnamefont {Chern}},
  \bibinfo {author} {\bibfnamefont {S.}~\bibnamefont {Kim}}, \bibinfo {author}
  {\bibfnamefont {P.~J.}\ \bibnamefont {Bereciartua}}, \bibinfo {author}
  {\bibfnamefont {S.}~\bibnamefont {Francoual}}, \bibinfo {author}
  {\bibfnamefont {Y.~B.}\ \bibnamefont {Kim}},\ and\ \bibinfo {author}
  {\bibfnamefont {Y.-J.}\ \bibnamefont {Kim}},\ }\bibfield  {title} {\bibinfo
  {title} {Ferromagnetic {K}itaev interaction and the origin of large magnetic
  anisotropy in $\alpha$-{R}u{C}l$_3$},\ }\href@noop {} {\bibfield  {journal}
  {\bibinfo  {journal} {Nature physics}\ }\textbf {\bibinfo {volume} {16}},\
  \bibinfo {pages} {837} (\bibinfo {year} {2020})}\BibitemShut {NoStop}%
\bibitem [{\citenamefont {Versteeg}\ \emph {et~al.}(2020)\citenamefont
  {Versteeg}, \citenamefont {Chiocchetta}, \citenamefont {Sekiguchi},
  \citenamefont {Aldea}, \citenamefont {Sahasrabudhe}, \citenamefont
  {Budzinauskas}, \citenamefont {Wang}, \citenamefont {Tsurkan}, \citenamefont
  {Loidl}, \citenamefont {Khomskii} \emph
  {et~al.}}]{versteeg2020nonequilibrium}%
  \BibitemOpen
  \bibfield  {author} {\bibinfo {author} {\bibfnamefont {R.}~\bibnamefont
  {Versteeg}}, \bibinfo {author} {\bibfnamefont {A.}~\bibnamefont
  {Chiocchetta}}, \bibinfo {author} {\bibfnamefont {F.}~\bibnamefont
  {Sekiguchi}}, \bibinfo {author} {\bibfnamefont {A.}~\bibnamefont {Aldea}},
  \bibinfo {author} {\bibfnamefont {A.}~\bibnamefont {Sahasrabudhe}}, \bibinfo
  {author} {\bibfnamefont {K.}~\bibnamefont {Budzinauskas}}, \bibinfo {author}
  {\bibfnamefont {Z.}~\bibnamefont {Wang}}, \bibinfo {author} {\bibfnamefont
  {V.}~\bibnamefont {Tsurkan}}, \bibinfo {author} {\bibfnamefont
  {A.}~\bibnamefont {Loidl}}, \bibinfo {author} {\bibfnamefont
  {D.}~\bibnamefont {Khomskii}}, \emph {et~al.},\ }\href@noop {} {\bibinfo
  {title} {Nonequilibrium quasistationary spin disordered state in the
  {K}itaev-{H}eisenberg magnet $\alpha$-{R}u{C}l$_3$}} (\bibinfo {year}
  {2020}),\ \Eprint {https://arxiv.org/abs/2005.14189} {arXiv:2005.14189
  [cond-mat.str-el]} \BibitemShut {NoStop}%
\bibitem [{\citenamefont {Suzuki}\ \emph {et~al.}(2021)\citenamefont {Suzuki},
  \citenamefont {Liu}, \citenamefont {Bertinshaw}, \citenamefont {Ueda},
  \citenamefont {Kim}, \citenamefont {Laha}, \citenamefont {Weber},
  \citenamefont {Yang}, \citenamefont {Wang}, \citenamefont {Takahashi} \emph
  {et~al.}}]{suzuki2021proximate}%
  \BibitemOpen
  \bibfield  {author} {\bibinfo {author} {\bibfnamefont {H.}~\bibnamefont
  {Suzuki}}, \bibinfo {author} {\bibfnamefont {H.}~\bibnamefont {Liu}},
  \bibinfo {author} {\bibfnamefont {J.}~\bibnamefont {Bertinshaw}}, \bibinfo
  {author} {\bibfnamefont {K.}~\bibnamefont {Ueda}}, \bibinfo {author}
  {\bibfnamefont {H.}~\bibnamefont {Kim}}, \bibinfo {author} {\bibfnamefont
  {S.}~\bibnamefont {Laha}}, \bibinfo {author} {\bibfnamefont {D.}~\bibnamefont
  {Weber}}, \bibinfo {author} {\bibfnamefont {Z.}~\bibnamefont {Yang}},
  \bibinfo {author} {\bibfnamefont {L.}~\bibnamefont {Wang}}, \bibinfo {author}
  {\bibfnamefont {H.}~\bibnamefont {Takahashi}}, \emph {et~al.},\ }\bibfield
  {title} {\bibinfo {title} {Proximate ferromagnetic state in the {K}itaev
  model material $\alpha$-{R}u{C}l$_3$},\ }\href@noop {} {\bibfield  {journal}
  {\bibinfo  {journal} {Nature Communications}\ }\textbf {\bibinfo {volume}
  {12}},\ \bibinfo {pages} {1} (\bibinfo {year} {2021})}\BibitemShut {NoStop}%
\bibitem [{\citenamefont {Janssen}\ and\ \citenamefont
  {Vojta}(2019)}]{janssen2019heisenberg}%
  \BibitemOpen
  \bibfield  {author} {\bibinfo {author} {\bibfnamefont {L.}~\bibnamefont
  {Janssen}}\ and\ \bibinfo {author} {\bibfnamefont {M.}~\bibnamefont
  {Vojta}},\ }\bibfield  {title} {\bibinfo {title} {Heisenberg-{K}itaev physics
  in magnetic fields},\ }\href@noop {} {\bibfield  {journal} {\bibinfo
  {journal} {Journal of Physics: Condensed Matter}\ }\textbf {\bibinfo {volume}
  {31}},\ \bibinfo {pages} {423002} (\bibinfo {year} {2019})}\BibitemShut
  {NoStop}%
\bibitem [{\citenamefont {Bittner}\ \emph {et~al.}(2020)\citenamefont
  {Bittner}, \citenamefont {Gole{\v{z}}}, \citenamefont {Eckstein},\ and\
  \citenamefont {Werner}}]{bittner2020photoenhanced}%
  \BibitemOpen
  \bibfield  {author} {\bibinfo {author} {\bibfnamefont {N.}~\bibnamefont
  {Bittner}}, \bibinfo {author} {\bibfnamefont {D.}~\bibnamefont
  {Gole{\v{z}}}}, \bibinfo {author} {\bibfnamefont {M.}~\bibnamefont
  {Eckstein}},\ and\ \bibinfo {author} {\bibfnamefont {P.}~\bibnamefont
  {Werner}},\ }\bibfield  {title} {\bibinfo {title} {Photoenhanced excitonic
  correlations in a {M}ott insulator with nonlocal interactions},\ }\href@noop
  {} {\bibfield  {journal} {\bibinfo  {journal} {Physical Review B}\ }\textbf
  {\bibinfo {volume} {101}},\ \bibinfo {pages} {085127} (\bibinfo {year}
  {2020})}\BibitemShut {NoStop}%
\bibitem [{\citenamefont {Sandilands}\ \emph
  {et~al.}(2016{\natexlab{a}})\citenamefont {Sandilands}, \citenamefont {Tian},
  \citenamefont {Reijnders}, \citenamefont {Kim}, \citenamefont {Plumb},
  \citenamefont {Kim}, \citenamefont {Kee},\ and\ \citenamefont
  {Burch}}]{sandilands2016spin}%
  \BibitemOpen
  \bibfield  {author} {\bibinfo {author} {\bibfnamefont {L.~J.}\ \bibnamefont
  {Sandilands}}, \bibinfo {author} {\bibfnamefont {Y.}~\bibnamefont {Tian}},
  \bibinfo {author} {\bibfnamefont {A.~A.}\ \bibnamefont {Reijnders}}, \bibinfo
  {author} {\bibfnamefont {H.-S.}\ \bibnamefont {Kim}}, \bibinfo {author}
  {\bibfnamefont {K.~W.}\ \bibnamefont {Plumb}}, \bibinfo {author}
  {\bibfnamefont {Y.-J.}\ \bibnamefont {Kim}}, \bibinfo {author} {\bibfnamefont
  {H.-Y.}\ \bibnamefont {Kee}},\ and\ \bibinfo {author} {\bibfnamefont {K.~S.}\
  \bibnamefont {Burch}},\ }\bibfield  {title} {\bibinfo {title} {Spin-orbit
  excitations and electronic structure of the putative {K}itaev magnet
  $\alpha$-{R}u{C}l$_3$},\ }\href@noop {} {\bibfield  {journal} {\bibinfo
  {journal} {Physical Review B}\ }\textbf {\bibinfo {volume} {93}},\ \bibinfo
  {pages} {075144} (\bibinfo {year} {2016}{\natexlab{a}})}\BibitemShut
  {NoStop}%
\bibitem [{\citenamefont {Sandilands}\ \emph
  {et~al.}(2016{\natexlab{b}})\citenamefont {Sandilands}, \citenamefont {Sohn},
  \citenamefont {Park}, \citenamefont {Kim}, \citenamefont {Kim}, \citenamefont
  {Sears}, \citenamefont {Kim},\ and\ \citenamefont
  {Noh}}]{sandilands2016optical}%
  \BibitemOpen
  \bibfield  {author} {\bibinfo {author} {\bibfnamefont {L.~J.}\ \bibnamefont
  {Sandilands}}, \bibinfo {author} {\bibfnamefont {C.~H.}\ \bibnamefont
  {Sohn}}, \bibinfo {author} {\bibfnamefont {H.~J.}\ \bibnamefont {Park}},
  \bibinfo {author} {\bibfnamefont {S.~Y.}\ \bibnamefont {Kim}}, \bibinfo
  {author} {\bibfnamefont {K.~W.}\ \bibnamefont {Kim}}, \bibinfo {author}
  {\bibfnamefont {J.~A.}\ \bibnamefont {Sears}}, \bibinfo {author}
  {\bibfnamefont {Y.-J.}\ \bibnamefont {Kim}},\ and\ \bibinfo {author}
  {\bibfnamefont {T.~W.}\ \bibnamefont {Noh}},\ }\bibfield  {title} {\bibinfo
  {title} {Optical probe of {H}eisenberg-{K}itaev magnetism in
  $\alpha$-{R}u{C}l$_3$},\ }\href@noop {} {\bibfield  {journal} {\bibinfo
  {journal} {Physical Review B}\ }\textbf {\bibinfo {volume} {94}},\ \bibinfo
  {pages} {195156} (\bibinfo {year} {2016}{\natexlab{b}})}\BibitemShut
  {NoStop}%
\bibitem [{\citenamefont {Beaurepaire}\ \emph {et~al.}(1996)\citenamefont
  {Beaurepaire}, \citenamefont {Merle}, \citenamefont {Daunois},\ and\
  \citenamefont {Bigot}}]{beaurepaire1996ultrafast}%
  \BibitemOpen
  \bibfield  {author} {\bibinfo {author} {\bibfnamefont {E.}~\bibnamefont
  {Beaurepaire}}, \bibinfo {author} {\bibfnamefont {J.-C.}\ \bibnamefont
  {Merle}}, \bibinfo {author} {\bibfnamefont {A.}~\bibnamefont {Daunois}},\
  and\ \bibinfo {author} {\bibfnamefont {J.-Y.}\ \bibnamefont {Bigot}},\
  }\bibfield  {title} {\bibinfo {title} {Ultrafast spin dynamics in
  ferromagnetic nickel},\ }\href@noop {} {\bibfield  {journal} {\bibinfo
  {journal} {Physical Review Letters}\ }\textbf {\bibinfo {volume} {76}},\
  \bibinfo {pages} {4250} (\bibinfo {year} {1996})}\BibitemShut {NoStop}%
\bibitem [{\citenamefont {Carpene}\ \emph {et~al.}(2015)\citenamefont
  {Carpene}, \citenamefont {Hedayat}, \citenamefont {Boschini},\ and\
  \citenamefont {Dallera}}]{carpene2015ultrafast}%
  \BibitemOpen
  \bibfield  {author} {\bibinfo {author} {\bibfnamefont {E.}~\bibnamefont
  {Carpene}}, \bibinfo {author} {\bibfnamefont {H.}~\bibnamefont {Hedayat}},
  \bibinfo {author} {\bibfnamefont {F.}~\bibnamefont {Boschini}},\ and\
  \bibinfo {author} {\bibfnamefont {C.}~\bibnamefont {Dallera}},\ }\bibfield
  {title} {\bibinfo {title} {Ultrafast demagnetization of metals: {C}ollapsed
  exchange versus collective excitations},\ }\href@noop {} {\bibfield
  {journal} {\bibinfo  {journal} {Physical Review B}\ }\textbf {\bibinfo
  {volume} {91}},\ \bibinfo {pages} {174414} (\bibinfo {year}
  {2015})}\BibitemShut {NoStop}%
\bibitem [{\citenamefont {Sirica}\ \emph {et~al.}(2021)\citenamefont {Sirica},
  \citenamefont {Hedayat}, \citenamefont {Bugini}, \citenamefont {Koehler},
  \citenamefont {Li}, \citenamefont {Parker}, \citenamefont {Mandrus},
  \citenamefont {Dallera}, \citenamefont {Carpene},\ and\ \citenamefont
  {Mannella}}]{sirica2021disentangling}%
  \BibitemOpen
  \bibfield  {author} {\bibinfo {author} {\bibfnamefont {N.}~\bibnamefont
  {Sirica}}, \bibinfo {author} {\bibfnamefont {H.}~\bibnamefont {Hedayat}},
  \bibinfo {author} {\bibfnamefont {D.}~\bibnamefont {Bugini}}, \bibinfo
  {author} {\bibfnamefont {M.}~\bibnamefont {Koehler}}, \bibinfo {author}
  {\bibfnamefont {L.}~\bibnamefont {Li}}, \bibinfo {author} {\bibfnamefont
  {D.}~\bibnamefont {Parker}}, \bibinfo {author} {\bibfnamefont
  {D.}~\bibnamefont {Mandrus}}, \bibinfo {author} {\bibfnamefont
  {C.}~\bibnamefont {Dallera}}, \bibinfo {author} {\bibfnamefont
  {E.}~\bibnamefont {Carpene}},\ and\ \bibinfo {author} {\bibfnamefont
  {N.}~\bibnamefont {Mannella}},\ }\bibfield  {title} {\bibinfo {title}
  {Disentangling electronic, lattice, and spin dynamics in the chiral
  helimagnet {Cr}$_\frac{1}{3}${N}b{S}$_2$},\ }\href@noop {} {\bibfield
  {journal} {\bibinfo  {journal} {Physical Review B}\ }\textbf {\bibinfo
  {volume} {104}},\ \bibinfo {pages} {174426} (\bibinfo {year}
  {2021})}\BibitemShut {NoStop}%
\bibitem [{\citenamefont {Lovinger}\ \emph {et~al.}(2020)\citenamefont
  {Lovinger}, \citenamefont {Brahlek}, \citenamefont {Kissin}, \citenamefont
  {Kennes}, \citenamefont {Millis}, \citenamefont {Engel-Herbert},\ and\
  \citenamefont {Averitt}}]{lovinger2020influence}%
  \BibitemOpen
  \bibfield  {author} {\bibinfo {author} {\bibfnamefont {D.}~\bibnamefont
  {Lovinger}}, \bibinfo {author} {\bibfnamefont {M.}~\bibnamefont {Brahlek}},
  \bibinfo {author} {\bibfnamefont {P.}~\bibnamefont {Kissin}}, \bibinfo
  {author} {\bibfnamefont {D.}~\bibnamefont {Kennes}}, \bibinfo {author}
  {\bibfnamefont {A.}~\bibnamefont {Millis}}, \bibinfo {author} {\bibfnamefont
  {R.}~\bibnamefont {Engel-Herbert}},\ and\ \bibinfo {author} {\bibfnamefont
  {R.}~\bibnamefont {Averitt}},\ }\bibfield  {title} {\bibinfo {title}
  {Influence of spin and orbital fluctuations on {M}ott-{H}ubbard exciton
  dynamics in {L}a{V}{O}$_3$ thin films},\ }\href@noop {} {\bibfield  {journal}
  {\bibinfo  {journal} {Physical Review B}\ }\textbf {\bibinfo {volume}
  {102}},\ \bibinfo {pages} {115143} (\bibinfo {year} {2020})}\BibitemShut
  {NoStop}%
\bibitem [{\citenamefont {Dolgirev}\ \emph {et~al.}(2020)\citenamefont
  {Dolgirev}, \citenamefont {Michael}, \citenamefont {Zong}, \citenamefont
  {Gedik},\ and\ \citenamefont {Demler}}]{dolgirev2020self}%
  \BibitemOpen
  \bibfield  {author} {\bibinfo {author} {\bibfnamefont {P.~E.}\ \bibnamefont
  {Dolgirev}}, \bibinfo {author} {\bibfnamefont {M.~H.}\ \bibnamefont
  {Michael}}, \bibinfo {author} {\bibfnamefont {A.}~\bibnamefont {Zong}},
  \bibinfo {author} {\bibfnamefont {N.}~\bibnamefont {Gedik}},\ and\ \bibinfo
  {author} {\bibfnamefont {E.}~\bibnamefont {Demler}},\ }\bibfield  {title}
  {\bibinfo {title} {Self-similar dynamics of order parameter fluctuations in
  pump-probe experiments},\ }\href@noop {} {\bibfield  {journal} {\bibinfo
  {journal} {Physical Review B}\ }\textbf {\bibinfo {volume} {101}},\ \bibinfo
  {pages} {174306} (\bibinfo {year} {2020})}\BibitemShut {NoStop}%
\bibitem [{\citenamefont {Afanasiev}\ \emph {et~al.}(2019)\citenamefont
  {Afanasiev}, \citenamefont {Gatilova}, \citenamefont {Groenendijk},
  \citenamefont {Ivanov}, \citenamefont {Gibert}, \citenamefont {Gariglio},
  \citenamefont {Mentink}, \citenamefont {Li}, \citenamefont {Dasari},
  \citenamefont {Eckstein} \emph {et~al.}}]{afanasiev2019ultrafast}%
  \BibitemOpen
  \bibfield  {author} {\bibinfo {author} {\bibfnamefont {D.}~\bibnamefont
  {Afanasiev}}, \bibinfo {author} {\bibfnamefont {A.}~\bibnamefont {Gatilova}},
  \bibinfo {author} {\bibfnamefont {D.}~\bibnamefont {Groenendijk}}, \bibinfo
  {author} {\bibfnamefont {B.}~\bibnamefont {Ivanov}}, \bibinfo {author}
  {\bibfnamefont {M.}~\bibnamefont {Gibert}}, \bibinfo {author} {\bibfnamefont
  {S.}~\bibnamefont {Gariglio}}, \bibinfo {author} {\bibfnamefont
  {J.}~\bibnamefont {Mentink}}, \bibinfo {author} {\bibfnamefont
  {J.}~\bibnamefont {Li}}, \bibinfo {author} {\bibfnamefont {N.}~\bibnamefont
  {Dasari}}, \bibinfo {author} {\bibfnamefont {M.}~\bibnamefont {Eckstein}},
  \emph {et~al.},\ }\bibfield  {title} {\bibinfo {title} {Ultrafast spin
  dynamics in photodoped spin-orbit {M}ott insulator {S}r$_2${I}r{O}$_4$},\
  }\href@noop {} {\bibfield  {journal} {\bibinfo  {journal} {Physical Review
  X}\ }\textbf {\bibinfo {volume} {9}},\ \bibinfo {pages} {021020} (\bibinfo
  {year} {2019})}\BibitemShut {NoStop}%
\bibitem [{\citenamefont {Lenar{\v{c}}i{\v{c}}}\ and\ \citenamefont
  {Prelov{\v{s}}ek}(2013)}]{lenarvcivc2013ultrafast}%
  \BibitemOpen
  \bibfield  {author} {\bibinfo {author} {\bibfnamefont {Z.}~\bibnamefont
  {Lenar{\v{c}}i{\v{c}}}}\ and\ \bibinfo {author} {\bibfnamefont
  {P.}~\bibnamefont {Prelov{\v{s}}ek}},\ }\bibfield  {title} {\bibinfo {title}
  {Ultrafast charge recombination in a photoexcited {M}ott-{H}ubbard
  insulator},\ }\href@noop {} {\bibfield  {journal} {\bibinfo  {journal}
  {Physical Review Letters}\ }\textbf {\bibinfo {volume} {111}},\ \bibinfo
  {pages} {016401} (\bibinfo {year} {2013})}\BibitemShut {NoStop}%
\bibitem [{\citenamefont {Yamakawa}\ \emph {et~al.}(2017)\citenamefont
  {Yamakawa}, \citenamefont {Miyamoto}, \citenamefont {Morimoto}, \citenamefont
  {Terashige}, \citenamefont {Yada}, \citenamefont {Kida}, \citenamefont
  {Suda}, \citenamefont {Yamamoto}, \citenamefont {Kato}, \citenamefont
  {Miyagawa} \emph {et~al.}}]{yamakawa2017mott}%
  \BibitemOpen
  \bibfield  {author} {\bibinfo {author} {\bibfnamefont {H.}~\bibnamefont
  {Yamakawa}}, \bibinfo {author} {\bibfnamefont {T.}~\bibnamefont {Miyamoto}},
  \bibinfo {author} {\bibfnamefont {T.}~\bibnamefont {Morimoto}}, \bibinfo
  {author} {\bibfnamefont {T.}~\bibnamefont {Terashige}}, \bibinfo {author}
  {\bibfnamefont {H.}~\bibnamefont {Yada}}, \bibinfo {author} {\bibfnamefont
  {N.}~\bibnamefont {Kida}}, \bibinfo {author} {\bibfnamefont {M.}~\bibnamefont
  {Suda}}, \bibinfo {author} {\bibfnamefont {H.}~\bibnamefont {Yamamoto}},
  \bibinfo {author} {\bibfnamefont {R.}~\bibnamefont {Kato}}, \bibinfo {author}
  {\bibfnamefont {K.}~\bibnamefont {Miyagawa}}, \emph {et~al.},\ }\bibfield
  {title} {\bibinfo {title} {Mott transition by an impulsive dielectric
  breakdown},\ }\href@noop {} {\bibfield  {journal} {\bibinfo  {journal}
  {Nature materials}\ }\textbf {\bibinfo {volume} {16}},\ \bibinfo {pages}
  {1100} (\bibinfo {year} {2017})}\BibitemShut {NoStop}%
\bibitem [{\citenamefont {Nevola}\ \emph {et~al.}(2021)\citenamefont {Nevola},
  \citenamefont {Bataller}, \citenamefont {Kumar}, \citenamefont {Sridhar},
  \citenamefont {Frick}, \citenamefont {O'Donnell}, \citenamefont {Ade},
  \citenamefont {Maggard}, \citenamefont {Kemper}, \citenamefont {Gundogdu}
  \emph {et~al.}}]{nevola2021timescales}%
  \BibitemOpen
  \bibfield  {author} {\bibinfo {author} {\bibfnamefont {D.}~\bibnamefont
  {Nevola}}, \bibinfo {author} {\bibfnamefont {A.}~\bibnamefont {Bataller}},
  \bibinfo {author} {\bibfnamefont {A.}~\bibnamefont {Kumar}}, \bibinfo
  {author} {\bibfnamefont {S.}~\bibnamefont {Sridhar}}, \bibinfo {author}
  {\bibfnamefont {J.}~\bibnamefont {Frick}}, \bibinfo {author} {\bibfnamefont
  {S.}~\bibnamefont {O'Donnell}}, \bibinfo {author} {\bibfnamefont
  {H.}~\bibnamefont {Ade}}, \bibinfo {author} {\bibfnamefont {P.~A.}\
  \bibnamefont {Maggard}}, \bibinfo {author} {\bibfnamefont {A.~F.}\
  \bibnamefont {Kemper}}, \bibinfo {author} {\bibfnamefont {K.}~\bibnamefont
  {Gundogdu}}, \emph {et~al.},\ }\bibfield  {title} {\bibinfo {title}
  {Timescales of excited state relaxation in $\alpha$-{R}u{C}l$_3$ observed by
  time-resolved two-photon photoemission spectroscopy},\ }\href@noop {}
  {\bibfield  {journal} {\bibinfo  {journal} {Physical Review B}\ }\textbf
  {\bibinfo {volume} {103}},\ \bibinfo {pages} {245105} (\bibinfo {year}
  {2021})}\BibitemShut {NoStop}%
\bibitem [{\citenamefont {Ferr{\'e}}\ and\ \citenamefont
  {Gehring}(1984)}]{ferre1984linear}%
  \BibitemOpen
  \bibfield  {author} {\bibinfo {author} {\bibfnamefont {J.}~\bibnamefont
  {Ferr{\'e}}}\ and\ \bibinfo {author} {\bibfnamefont {G.}~\bibnamefont
  {Gehring}},\ }\bibfield  {title} {\bibinfo {title} {Linear optical
  birefringence of magnetic crystals},\ }\href@noop {} {\bibfield  {journal}
  {\bibinfo  {journal} {Reports on Progress in Physics}\ }\textbf {\bibinfo
  {volume} {47}},\ \bibinfo {pages} {513} (\bibinfo {year} {1984})}\BibitemShut
  {NoStop}%
\bibitem [{\citenamefont {Mak}\ \emph {et~al.}(2019)\citenamefont {Mak},
  \citenamefont {Shan},\ and\ \citenamefont {Ralph}}]{mak2019probing}%
  \BibitemOpen
  \bibfield  {author} {\bibinfo {author} {\bibfnamefont {K.~F.}\ \bibnamefont
  {Mak}}, \bibinfo {author} {\bibfnamefont {J.}~\bibnamefont {Shan}},\ and\
  \bibinfo {author} {\bibfnamefont {D.~C.}\ \bibnamefont {Ralph}},\ }\bibfield
  {title} {\bibinfo {title} {Probing and controlling magnetic states in 2{D}
  layered magnetic materials},\ }\href@noop {} {\bibfield  {journal} {\bibinfo
  {journal} {Nature Reviews Physics}\ }\textbf {\bibinfo {volume} {1}},\
  \bibinfo {pages} {646} (\bibinfo {year} {2019})}\BibitemShut {NoStop}%
\bibitem [{\citenamefont {Saidl}\ \emph {et~al.}(2017)\citenamefont {Saidl},
  \citenamefont {N{\v{e}}mec}, \citenamefont {Wadley}, \citenamefont {Hills},
  \citenamefont {Campion}, \citenamefont {Nov{\'a}k}, \citenamefont {Edmonds},
  \citenamefont {Maccherozzi}, \citenamefont {Dhesi}, \citenamefont {Gallagher}
  \emph {et~al.}}]{saidl2017optical}%
  \BibitemOpen
  \bibfield  {author} {\bibinfo {author} {\bibfnamefont {V.}~\bibnamefont
  {Saidl}}, \bibinfo {author} {\bibfnamefont {P.}~\bibnamefont {N{\v{e}}mec}},
  \bibinfo {author} {\bibfnamefont {P.}~\bibnamefont {Wadley}}, \bibinfo
  {author} {\bibfnamefont {V.}~\bibnamefont {Hills}}, \bibinfo {author}
  {\bibfnamefont {R.}~\bibnamefont {Campion}}, \bibinfo {author} {\bibfnamefont
  {V.}~\bibnamefont {Nov{\'a}k}}, \bibinfo {author} {\bibfnamefont
  {K.}~\bibnamefont {Edmonds}}, \bibinfo {author} {\bibfnamefont
  {F.}~\bibnamefont {Maccherozzi}}, \bibinfo {author} {\bibfnamefont
  {S.}~\bibnamefont {Dhesi}}, \bibinfo {author} {\bibfnamefont
  {B.}~\bibnamefont {Gallagher}}, \emph {et~al.},\ }\bibfield  {title}
  {\bibinfo {title} {Optical determination of the {N}{\'e}el vector in a
  {C}u{M}n{A}s thin-film antiferromagnet},\ }\href@noop {} {\bibfield
  {journal} {\bibinfo  {journal} {Nature Photonics}\ }\textbf {\bibinfo
  {volume} {11}},\ \bibinfo {pages} {91} (\bibinfo {year} {2017})}\BibitemShut
  {NoStop}%
\bibitem [{\citenamefont {Zhang}\ \emph
  {et~al.}(2021{\natexlab{a}})\citenamefont {Zhang}, \citenamefont {Hwangbo},
  \citenamefont {Wang}, \citenamefont {Jiang}, \citenamefont {Chu},
  \citenamefont {Wen}, \citenamefont {Xiao},\ and\ \citenamefont
  {Xu}}]{zhang2021observation}%
  \BibitemOpen
  \bibfield  {author} {\bibinfo {author} {\bibfnamefont {Q.}~\bibnamefont
  {Zhang}}, \bibinfo {author} {\bibfnamefont {K.}~\bibnamefont {Hwangbo}},
  \bibinfo {author} {\bibfnamefont {C.}~\bibnamefont {Wang}}, \bibinfo {author}
  {\bibfnamefont {Q.}~\bibnamefont {Jiang}}, \bibinfo {author} {\bibfnamefont
  {J.-H.}\ \bibnamefont {Chu}}, \bibinfo {author} {\bibfnamefont
  {H.}~\bibnamefont {Wen}}, \bibinfo {author} {\bibfnamefont {D.}~\bibnamefont
  {Xiao}},\ and\ \bibinfo {author} {\bibfnamefont {X.}~\bibnamefont {Xu}},\
  }\bibfield  {title} {\bibinfo {title} {Observation of {G}iant {O}ptical
  {L}inear {D}ichroism in a {Z}igzag {A}ntiferromagnet fe{P}s$_3$},\
  }\href@noop {} {\bibfield  {journal} {\bibinfo  {journal} {Nano Letters}\
  }\textbf {\bibinfo {volume} {21}},\ \bibinfo {pages} {6938} (\bibinfo {year}
  {2021}{\natexlab{a}})}\BibitemShut {NoStop}%
\bibitem [{\citenamefont {Zhang}\ \emph
  {et~al.}(2021{\natexlab{b}})\citenamefont {Zhang}, \citenamefont {Jiang},
  \citenamefont {Lee}, \citenamefont {Lee}, \citenamefont {Mak},\ and\
  \citenamefont {Shan}}]{zhang2021spin}%
  \BibitemOpen
  \bibfield  {author} {\bibinfo {author} {\bibfnamefont {X.-X.}\ \bibnamefont
  {Zhang}}, \bibinfo {author} {\bibfnamefont {S.}~\bibnamefont {Jiang}},
  \bibinfo {author} {\bibfnamefont {J.}~\bibnamefont {Lee}}, \bibinfo {author}
  {\bibfnamefont {C.}~\bibnamefont {Lee}}, \bibinfo {author} {\bibfnamefont
  {K.~F.}\ \bibnamefont {Mak}},\ and\ \bibinfo {author} {\bibfnamefont
  {J.}~\bibnamefont {Shan}},\ }\bibfield  {title} {\bibinfo {title} {Spin
  dynamics slowdown near the antiferromagnetic critical point in atomically
  thin {F}e{P}{S}$_3$},\ }\href@noop {} {\bibfield  {journal} {\bibinfo
  {journal} {Nano Letters}\ } (\bibinfo {year}
  {2021}{\natexlab{b}})}\BibitemShut {NoStop}%
\bibitem [{\citenamefont {Sahasrabudhe}\ \emph {et~al.}(2020)\citenamefont
  {Sahasrabudhe}, \citenamefont {Kaib}, \citenamefont {Reschke}, \citenamefont
  {German}, \citenamefont {Koethe}, \citenamefont {Buhot}, \citenamefont
  {Kamenskyi}, \citenamefont {Hickey}, \citenamefont {Becker}, \citenamefont
  {Tsurkan} \emph {et~al.}}]{sahasrabudhe2020high}%
  \BibitemOpen
  \bibfield  {author} {\bibinfo {author} {\bibfnamefont {A.}~\bibnamefont
  {Sahasrabudhe}}, \bibinfo {author} {\bibfnamefont {D.}~\bibnamefont {Kaib}},
  \bibinfo {author} {\bibfnamefont {S.}~\bibnamefont {Reschke}}, \bibinfo
  {author} {\bibfnamefont {R.}~\bibnamefont {German}}, \bibinfo {author}
  {\bibfnamefont {T.}~\bibnamefont {Koethe}}, \bibinfo {author} {\bibfnamefont
  {J.}~\bibnamefont {Buhot}}, \bibinfo {author} {\bibfnamefont
  {D.}~\bibnamefont {Kamenskyi}}, \bibinfo {author} {\bibfnamefont
  {C.}~\bibnamefont {Hickey}}, \bibinfo {author} {\bibfnamefont
  {P.}~\bibnamefont {Becker}}, \bibinfo {author} {\bibfnamefont
  {V.}~\bibnamefont {Tsurkan}}, \emph {et~al.},\ }\bibfield  {title} {\bibinfo
  {title} {High-field quantum disordered state in $\alpha$-{R}u{C}l$_3$: {S}pin
  flips, bound states, and multiparticle continuum},\ }\href@noop {} {\bibfield
   {journal} {\bibinfo  {journal} {Physical Review B}\ }\textbf {\bibinfo
  {volume} {101}},\ \bibinfo {pages} {140410} (\bibinfo {year}
  {2020})}\BibitemShut {NoStop}%
\bibitem [{\citenamefont {Mi}\ \emph {et~al.}(2021)\citenamefont {Mi},
  \citenamefont {Wang}, \citenamefont {Gui}, \citenamefont {Pi}, \citenamefont
  {Zheng}, \citenamefont {Yang}, \citenamefont {Gan}, \citenamefont {Wang},
  \citenamefont {Li}, \citenamefont {Wang} \emph {et~al.}}]{mi2021stacking}%
  \BibitemOpen
  \bibfield  {author} {\bibinfo {author} {\bibfnamefont {X.}~\bibnamefont
  {Mi}}, \bibinfo {author} {\bibfnamefont {X.}~\bibnamefont {Wang}}, \bibinfo
  {author} {\bibfnamefont {H.}~\bibnamefont {Gui}}, \bibinfo {author}
  {\bibfnamefont {M.}~\bibnamefont {Pi}}, \bibinfo {author} {\bibfnamefont
  {T.}~\bibnamefont {Zheng}}, \bibinfo {author} {\bibfnamefont
  {K.}~\bibnamefont {Yang}}, \bibinfo {author} {\bibfnamefont {Y.}~\bibnamefont
  {Gan}}, \bibinfo {author} {\bibfnamefont {P.}~\bibnamefont {Wang}}, \bibinfo
  {author} {\bibfnamefont {A.}~\bibnamefont {Li}}, \bibinfo {author}
  {\bibfnamefont {A.}~\bibnamefont {Wang}}, \emph {et~al.},\ }\bibfield
  {title} {\bibinfo {title} {Stacking faults in $\alpha$-{R}u{C}l$_3$ revealed
  by local electric polarization},\ }\href@noop {} {\bibfield  {journal}
  {\bibinfo  {journal} {Physical Review B}\ }\textbf {\bibinfo {volume}
  {103}},\ \bibinfo {pages} {174413} (\bibinfo {year} {2021})}\BibitemShut
  {NoStop}%
\bibitem [{\citenamefont {Tesa{\v{r}}ov{\'a}}\ \emph
  {et~al.}(2014)\citenamefont {Tesa{\v{r}}ov{\'a}}, \citenamefont
  {Ostatnick{\`y}}, \citenamefont {Nov{\'a}k}, \citenamefont {Olejn{\'\i}k},
  \citenamefont {{\v{S}}ubrt}, \citenamefont {Reichlov{\'a}}, \citenamefont
  {Ellis}, \citenamefont {Mukherjee}, \citenamefont {Lee}, \citenamefont
  {Sipahi} \emph {et~al.}}]{tesavrova2014systematic}%
  \BibitemOpen
  \bibfield  {author} {\bibinfo {author} {\bibfnamefont {N.}~\bibnamefont
  {Tesa{\v{r}}ov{\'a}}}, \bibinfo {author} {\bibfnamefont {T.}~\bibnamefont
  {Ostatnick{\`y}}}, \bibinfo {author} {\bibfnamefont {V.}~\bibnamefont
  {Nov{\'a}k}}, \bibinfo {author} {\bibfnamefont {K.}~\bibnamefont
  {Olejn{\'\i}k}}, \bibinfo {author} {\bibfnamefont {J.}~\bibnamefont
  {{\v{S}}ubrt}}, \bibinfo {author} {\bibfnamefont {H.}~\bibnamefont
  {Reichlov{\'a}}}, \bibinfo {author} {\bibfnamefont {C.}~\bibnamefont
  {Ellis}}, \bibinfo {author} {\bibfnamefont {A.}~\bibnamefont {Mukherjee}},
  \bibinfo {author} {\bibfnamefont {J.}~\bibnamefont {Lee}}, \bibinfo {author}
  {\bibfnamefont {G.~M.}\ \bibnamefont {Sipahi}}, \emph {et~al.},\ }\bibfield
  {title} {\bibinfo {title} {Systematic study of magnetic linear dichroism and
  birefringence in ({G}a, {M}n) {A}s},\ }\href@noop {} {\bibfield  {journal}
  {\bibinfo  {journal} {Physical Review B}\ }\textbf {\bibinfo {volume} {89}},\
  \bibinfo {pages} {085203} (\bibinfo {year} {2014})}\BibitemShut {NoStop}%
\bibitem [{\citenamefont {Prasankumar}\ and\ \citenamefont
  {Taylor}(2016)}]{prasankumar2016optical}%
  \BibitemOpen
  \bibfield  {author} {\bibinfo {author} {\bibfnamefont {R.~P.}\ \bibnamefont
  {Prasankumar}}\ and\ \bibinfo {author} {\bibfnamefont {A.~J.}\ \bibnamefont
  {Taylor}},\ }\href@noop {} {\emph {\bibinfo {title} {Optical techniques for
  solid-state materials characterization}}}\ (\bibinfo  {publisher} {CRC
  press},\ \bibinfo {year} {2016})\ Chap.~\bibinfo {chapter} {13}\BibitemShut
  {NoStop}%
\bibitem [{\citenamefont {Ma}\ \emph {et~al.}(2018)\citenamefont {Ma},
  \citenamefont {Ran}, \citenamefont {Wang}, \citenamefont {Bao}, \citenamefont
  {Cai}, \citenamefont {Li},\ and\ \citenamefont {Wen}}]{ma2018recent}%
  \BibitemOpen
  \bibfield  {author} {\bibinfo {author} {\bibfnamefont {Z.}~\bibnamefont
  {Ma}}, \bibinfo {author} {\bibfnamefont {K.}~\bibnamefont {Ran}}, \bibinfo
  {author} {\bibfnamefont {J.}~\bibnamefont {Wang}}, \bibinfo {author}
  {\bibfnamefont {S.}~\bibnamefont {Bao}}, \bibinfo {author} {\bibfnamefont
  {Z.}~\bibnamefont {Cai}}, \bibinfo {author} {\bibfnamefont {S.}~\bibnamefont
  {Li}},\ and\ \bibinfo {author} {\bibfnamefont {J.}~\bibnamefont {Wen}},\
  }\bibfield  {title} {\bibinfo {title} {Recent progress on magnetic-field
  studies on quantum-spin-liquid candidates},\ }\href@noop {} {\bibfield
  {journal} {\bibinfo  {journal} {Chinese Physics B}\ }\textbf {\bibinfo
  {volume} {27}},\ \bibinfo {pages} {106101} (\bibinfo {year}
  {2018})}\BibitemShut {NoStop}%
\bibitem [{\citenamefont {Lenar{\v{c}}i{\v{c}}}\ and\ \citenamefont
  {Prelov{\v{s}}ek}(2014)}]{lenarvcivc2014charge}%
  \BibitemOpen
  \bibfield  {author} {\bibinfo {author} {\bibfnamefont {Z.}~\bibnamefont
  {Lenar{\v{c}}i{\v{c}}}}\ and\ \bibinfo {author} {\bibfnamefont
  {P.}~\bibnamefont {Prelov{\v{s}}ek}},\ }\bibfield  {title} {\bibinfo {title}
  {Charge recombination in undoped cuprates},\ }\href@noop {} {\bibfield
  {journal} {\bibinfo  {journal} {Physical Review B}\ }\textbf {\bibinfo
  {volume} {90}},\ \bibinfo {pages} {235136} (\bibinfo {year}
  {2014})}\BibitemShut {NoStop}%
\bibitem [{\citenamefont
  {Lenar{\v{c}}i{\v{c}}}(2015)}]{lenarvcivc2015nonequilibrium}%
  \BibitemOpen
  \bibfield  {author} {\bibinfo {author} {\bibfnamefont {Z.}~\bibnamefont
  {Lenar{\v{c}}i{\v{c}}}},\ }\emph {\bibinfo {title} {Nonequilibrium properties
  of Mott insulators: doctoral thesis}},\ \href@noop {} {Ph.D. thesis},\
  \bibinfo  {school} {Univerza v Ljubljani, Fakulteta za matematiko in fiziko}
  (\bibinfo {year} {2015})\BibitemShut {NoStop}%
\bibitem [{\citenamefont {Okazaki}\ \emph {et~al.}(2018)\citenamefont
  {Okazaki}, \citenamefont {Ogawa}, \citenamefont {Suzuki}, \citenamefont
  {Yamamoto}, \citenamefont {Someya}, \citenamefont {Michimae}, \citenamefont
  {Watanabe}, \citenamefont {Lu}, \citenamefont {Nohara}, \citenamefont
  {Takagi} \emph {et~al.}}]{okazaki2018photo}%
  \BibitemOpen
  \bibfield  {author} {\bibinfo {author} {\bibfnamefont {K.}~\bibnamefont
  {Okazaki}}, \bibinfo {author} {\bibfnamefont {Y.}~\bibnamefont {Ogawa}},
  \bibinfo {author} {\bibfnamefont {T.}~\bibnamefont {Suzuki}}, \bibinfo
  {author} {\bibfnamefont {T.}~\bibnamefont {Yamamoto}}, \bibinfo {author}
  {\bibfnamefont {T.}~\bibnamefont {Someya}}, \bibinfo {author} {\bibfnamefont
  {S.}~\bibnamefont {Michimae}}, \bibinfo {author} {\bibfnamefont
  {M.}~\bibnamefont {Watanabe}}, \bibinfo {author} {\bibfnamefont
  {Y.}~\bibnamefont {Lu}}, \bibinfo {author} {\bibfnamefont {M.}~\bibnamefont
  {Nohara}}, \bibinfo {author} {\bibfnamefont {H.}~\bibnamefont {Takagi}},
  \emph {et~al.},\ }\bibfield  {title} {\bibinfo {title} {Photo-induced
  semimetallic states realised in electron-hole coupled insulators},\
  }\href@noop {} {\bibfield  {journal} {\bibinfo  {journal} {Nature
  communications}\ }\textbf {\bibinfo {volume} {9}},\ \bibinfo {pages} {1}
  (\bibinfo {year} {2018})}\BibitemShut {NoStop}%
\bibitem [{\citenamefont {Rohwer}\ \emph {et~al.}(2011)\citenamefont {Rohwer},
  \citenamefont {Hellmann}, \citenamefont {Wiesenmayer}, \citenamefont {Sohrt},
  \citenamefont {Stange}, \citenamefont {Slomski}, \citenamefont {Carr},
  \citenamefont {Liu}, \citenamefont {Avila}, \citenamefont {Kall{\"a}ne} \emph
  {et~al.}}]{rohwer2011collapse}%
  \BibitemOpen
  \bibfield  {author} {\bibinfo {author} {\bibfnamefont {T.}~\bibnamefont
  {Rohwer}}, \bibinfo {author} {\bibfnamefont {S.}~\bibnamefont {Hellmann}},
  \bibinfo {author} {\bibfnamefont {M.}~\bibnamefont {Wiesenmayer}}, \bibinfo
  {author} {\bibfnamefont {C.}~\bibnamefont {Sohrt}}, \bibinfo {author}
  {\bibfnamefont {A.}~\bibnamefont {Stange}}, \bibinfo {author} {\bibfnamefont
  {B.}~\bibnamefont {Slomski}}, \bibinfo {author} {\bibfnamefont
  {A.}~\bibnamefont {Carr}}, \bibinfo {author} {\bibfnamefont {Y.}~\bibnamefont
  {Liu}}, \bibinfo {author} {\bibfnamefont {L.~M.}\ \bibnamefont {Avila}},
  \bibinfo {author} {\bibfnamefont {M.}~\bibnamefont {Kall{\"a}ne}}, \emph
  {et~al.},\ }\bibfield  {title} {\bibinfo {title} {Collapse of long-range
  charge order tracked by time-resolved photoemission at high momenta},\
  }\href@noop {} {\bibfield  {journal} {\bibinfo  {journal} {Nature}\ }\textbf
  {\bibinfo {volume} {471}},\ \bibinfo {pages} {490} (\bibinfo {year}
  {2011})}\BibitemShut {NoStop}%
\bibitem [{\citenamefont {Hedayat}\ \emph {et~al.}(2019)\citenamefont
  {Hedayat}, \citenamefont {Sayers}, \citenamefont {Bugini}, \citenamefont
  {Dallera}, \citenamefont {Wolverson}, \citenamefont {Batten}, \citenamefont
  {Karbassi}, \citenamefont {Friedemann}, \citenamefont {Cerullo},
  \citenamefont {van Wezel} \emph {et~al.}}]{hedayat2019excitonic}%
  \BibitemOpen
  \bibfield  {author} {\bibinfo {author} {\bibfnamefont {H.}~\bibnamefont
  {Hedayat}}, \bibinfo {author} {\bibfnamefont {C.~J.}\ \bibnamefont {Sayers}},
  \bibinfo {author} {\bibfnamefont {D.}~\bibnamefont {Bugini}}, \bibinfo
  {author} {\bibfnamefont {C.}~\bibnamefont {Dallera}}, \bibinfo {author}
  {\bibfnamefont {D.}~\bibnamefont {Wolverson}}, \bibinfo {author}
  {\bibfnamefont {T.}~\bibnamefont {Batten}}, \bibinfo {author} {\bibfnamefont
  {S.}~\bibnamefont {Karbassi}}, \bibinfo {author} {\bibfnamefont
  {S.}~\bibnamefont {Friedemann}}, \bibinfo {author} {\bibfnamefont
  {G.}~\bibnamefont {Cerullo}}, \bibinfo {author} {\bibfnamefont
  {J.}~\bibnamefont {van Wezel}}, \emph {et~al.},\ }\bibfield  {title}
  {\bibinfo {title} {Excitonic and lattice contributions to the charge density
  wave in 1 {T}-{T}i{S}e$_2$ revealed by a phonon bottleneck},\ }\href@noop {}
  {\bibfield  {journal} {\bibinfo  {journal} {Physical Review Research}\
  }\textbf {\bibinfo {volume} {1}},\ \bibinfo {pages} {023029} (\bibinfo {year}
  {2019})}\BibitemShut {NoStop}%
\bibitem [{\citenamefont {Wagner}\ \emph {et~al.}(2014)\citenamefont {Wagner},
  \citenamefont {McLeod}, \citenamefont {Maddox}, \citenamefont {Fei},
  \citenamefont {Liu}, \citenamefont {Averitt}, \citenamefont {Fogler},
  \citenamefont {Bank}, \citenamefont {Keilmann},\ and\ \citenamefont
  {Basov}}]{wagner2014ultrafast}%
  \BibitemOpen
  \bibfield  {author} {\bibinfo {author} {\bibfnamefont {M.}~\bibnamefont
  {Wagner}}, \bibinfo {author} {\bibfnamefont {A.~S.}\ \bibnamefont {McLeod}},
  \bibinfo {author} {\bibfnamefont {S.~J.}\ \bibnamefont {Maddox}}, \bibinfo
  {author} {\bibfnamefont {Z.}~\bibnamefont {Fei}}, \bibinfo {author}
  {\bibfnamefont {M.}~\bibnamefont {Liu}}, \bibinfo {author} {\bibfnamefont
  {R.~D.}\ \bibnamefont {Averitt}}, \bibinfo {author} {\bibfnamefont {M.~M.}\
  \bibnamefont {Fogler}}, \bibinfo {author} {\bibfnamefont {S.~R.}\
  \bibnamefont {Bank}}, \bibinfo {author} {\bibfnamefont {F.}~\bibnamefont
  {Keilmann}},\ and\ \bibinfo {author} {\bibfnamefont {D.}~\bibnamefont
  {Basov}},\ }\bibfield  {title} {\bibinfo {title} {Ultrafast dynamics of
  surface plasmons in {I}n{A}s by time-resolved infrared nanospectroscopy},\
  }\href@noop {} {\bibfield  {journal} {\bibinfo  {journal} {Nano letters}\
  }\textbf {\bibinfo {volume} {14}},\ \bibinfo {pages} {4529} (\bibinfo {year}
  {2014})}\BibitemShut {NoStop}%
\bibitem [{\citenamefont {Radu}\ \emph {et~al.}(2010)\citenamefont {Radu},
  \citenamefont {Stamm}, \citenamefont {Pontius}, \citenamefont {Kachel},
  \citenamefont {Ramm}, \citenamefont {Thiele}, \citenamefont {D{\"u}rr},\ and\
  \citenamefont {Back}}]{radu2010laser}%
  \BibitemOpen
  \bibfield  {author} {\bibinfo {author} {\bibfnamefont {I.}~\bibnamefont
  {Radu}}, \bibinfo {author} {\bibfnamefont {C.}~\bibnamefont {Stamm}},
  \bibinfo {author} {\bibfnamefont {N.}~\bibnamefont {Pontius}}, \bibinfo
  {author} {\bibfnamefont {T.}~\bibnamefont {Kachel}}, \bibinfo {author}
  {\bibfnamefont {P.}~\bibnamefont {Ramm}}, \bibinfo {author} {\bibfnamefont
  {J.-U.}\ \bibnamefont {Thiele}}, \bibinfo {author} {\bibfnamefont
  {H.}~\bibnamefont {D{\"u}rr}},\ and\ \bibinfo {author} {\bibfnamefont
  {C.}~\bibnamefont {Back}},\ }\bibfield  {title} {\bibinfo {title}
  {Laser-induced generation and quenching of magnetization on {F}e{R}h studied
  with time-resolved x-ray magnetic circular dichroism},\ }\href@noop {}
  {\bibfield  {journal} {\bibinfo  {journal} {Physical Review B}\ }\textbf
  {\bibinfo {volume} {81}},\ \bibinfo {pages} {104415} (\bibinfo {year}
  {2010})}\BibitemShut {NoStop}%
\bibitem [{\citenamefont {Afanasiev}\ \emph {et~al.}(2016)\citenamefont
  {Afanasiev}, \citenamefont {Ivanov}, \citenamefont {Kirilyuk}, \citenamefont
  {Rasing}, \citenamefont {Pisarev},\ and\ \citenamefont
  {Kimel}}]{afanasiev2016control}%
  \BibitemOpen
  \bibfield  {author} {\bibinfo {author} {\bibfnamefont {D.}~\bibnamefont
  {Afanasiev}}, \bibinfo {author} {\bibfnamefont {B.}~\bibnamefont {Ivanov}},
  \bibinfo {author} {\bibfnamefont {A.}~\bibnamefont {Kirilyuk}}, \bibinfo
  {author} {\bibfnamefont {T.}~\bibnamefont {Rasing}}, \bibinfo {author}
  {\bibfnamefont {R.}~\bibnamefont {Pisarev}},\ and\ \bibinfo {author}
  {\bibfnamefont {A.}~\bibnamefont {Kimel}},\ }\bibfield  {title} {\bibinfo
  {title} {Control of the ultrafast photoinduced magnetization across the
  {M}orin transition in {D}y{F}e{O}$_3$},\ }\href@noop {} {\bibfield  {journal}
  {\bibinfo  {journal} {Physical Review Letters}\ }\textbf {\bibinfo {volume}
  {116}},\ \bibinfo {pages} {097401} (\bibinfo {year} {2016})}\BibitemShut
  {NoStop}%
\bibitem [{\citenamefont {Bergman}\ \emph {et~al.}(2006)\citenamefont
  {Bergman}, \citenamefont {Ju}, \citenamefont {Hohlfeld}, \citenamefont
  {van~de Veerdonk}, \citenamefont {Kim}, \citenamefont {Wu}, \citenamefont
  {Weller},\ and\ \citenamefont {Koopmans}}]{bergman2006identifying}%
  \BibitemOpen
  \bibfield  {author} {\bibinfo {author} {\bibfnamefont {B.}~\bibnamefont
  {Bergman}}, \bibinfo {author} {\bibfnamefont {G.}~\bibnamefont {Ju}},
  \bibinfo {author} {\bibfnamefont {J.}~\bibnamefont {Hohlfeld}}, \bibinfo
  {author} {\bibfnamefont {R.~J.}\ \bibnamefont {van~de Veerdonk}}, \bibinfo
  {author} {\bibfnamefont {J.-Y.}\ \bibnamefont {Kim}}, \bibinfo {author}
  {\bibfnamefont {X.}~\bibnamefont {Wu}}, \bibinfo {author} {\bibfnamefont
  {D.}~\bibnamefont {Weller}},\ and\ \bibinfo {author} {\bibfnamefont
  {B.}~\bibnamefont {Koopmans}},\ }\bibfield  {title} {\bibinfo {title}
  {Identifying growth mechanisms for laser-induced magnetization in {F}e{R}h},\
  }\href@noop {} {\bibfield  {journal} {\bibinfo  {journal} {Physical Review
  B}\ }\textbf {\bibinfo {volume} {73}},\ \bibinfo {pages} {060407} (\bibinfo
  {year} {2006})}\BibitemShut {NoStop}%
\bibitem [{\citenamefont {Koopmans}\ \emph {et~al.}(2010)\citenamefont
  {Koopmans}, \citenamefont {Malinowski}, \citenamefont {Dalla~Longa},
  \citenamefont {Steiauf}, \citenamefont {F{\"a}hnle}, \citenamefont {Roth},
  \citenamefont {Cinchetti},\ and\ \citenamefont
  {Aeschlimann}}]{koopmans2010explaining}%
  \BibitemOpen
  \bibfield  {author} {\bibinfo {author} {\bibfnamefont {B.}~\bibnamefont
  {Koopmans}}, \bibinfo {author} {\bibfnamefont {G.}~\bibnamefont
  {Malinowski}}, \bibinfo {author} {\bibfnamefont {F.}~\bibnamefont
  {Dalla~Longa}}, \bibinfo {author} {\bibfnamefont {D.}~\bibnamefont
  {Steiauf}}, \bibinfo {author} {\bibfnamefont {M.}~\bibnamefont {F{\"a}hnle}},
  \bibinfo {author} {\bibfnamefont {T.}~\bibnamefont {Roth}}, \bibinfo {author}
  {\bibfnamefont {M.}~\bibnamefont {Cinchetti}},\ and\ \bibinfo {author}
  {\bibfnamefont {M.}~\bibnamefont {Aeschlimann}},\ }\bibfield  {title}
  {\bibinfo {title} {Explaining the paradoxical diversity of ultrafast
  laser-induced demagnetization},\ }\href@noop {} {\bibfield  {journal}
  {\bibinfo  {journal} {Nature materials}\ }\textbf {\bibinfo {volume} {9}},\
  \bibinfo {pages} {259} (\bibinfo {year} {2010})}\BibitemShut {NoStop}%
\bibitem [{\citenamefont {Pastor}\ \emph {et~al.}(2021)\citenamefont {Pastor},
  \citenamefont {Moreno-Menc{\'\i}a}, \citenamefont {Monti}, \citenamefont
  {Johnson}, \citenamefont {Fleischmann}, \citenamefont {Wang}, \citenamefont
  {Shi}, \citenamefont {Liu}, \citenamefont {Mazzone}, \citenamefont {Dean}
  \emph {et~al.}}]{pastor2021non}%
  \BibitemOpen
  \bibfield  {author} {\bibinfo {author} {\bibfnamefont {E.}~\bibnamefont
  {Pastor}}, \bibinfo {author} {\bibfnamefont {D.}~\bibnamefont
  {Moreno-Menc{\'\i}a}}, \bibinfo {author} {\bibfnamefont {M.}~\bibnamefont
  {Monti}}, \bibinfo {author} {\bibfnamefont {A.~S.}\ \bibnamefont {Johnson}},
  \bibinfo {author} {\bibfnamefont {N.}~\bibnamefont {Fleischmann}}, \bibinfo
  {author} {\bibfnamefont {C.}~\bibnamefont {Wang}}, \bibinfo {author}
  {\bibfnamefont {Y.}~\bibnamefont {Shi}}, \bibinfo {author} {\bibfnamefont
  {X.}~\bibnamefont {Liu}}, \bibinfo {author} {\bibfnamefont {D.~G.}\
  \bibnamefont {Mazzone}}, \bibinfo {author} {\bibfnamefont {M.~P.}\
  \bibnamefont {Dean}}, \emph {et~al.},\ }\bibfield  {title} {\bibinfo {title}
  {Non-thermal breaking of magnetic order via photo-generated spin defects},\
  }\href@noop {} {\bibfield  {journal} {\bibinfo  {journal} {arXiv preprint
  arXiv:2104.04294}\ } (\bibinfo {year} {2021})}\BibitemShut {NoStop}%
\bibitem [{\citenamefont {Li}\ \emph {et~al.}(2020)\citenamefont {Li},
  \citenamefont {Dasari},\ and\ \citenamefont {Eckstein}}]{li2020ultrafast}%
  \BibitemOpen
  \bibfield  {author} {\bibinfo {author} {\bibfnamefont {J.}~\bibnamefont
  {Li}}, \bibinfo {author} {\bibfnamefont {N.}~\bibnamefont {Dasari}},\ and\
  \bibinfo {author} {\bibfnamefont {M.}~\bibnamefont {Eckstein}},\ }\bibfield
  {title} {\bibinfo {title} {Ultrafast dynamics in relativistic {M}ott
  insulators},\ }\href@noop {} {\bibfield  {journal} {\bibinfo  {journal}
  {arXiv preprint arXiv:2010.09253}\ } (\bibinfo {year} {2020})}\BibitemShut
  {NoStop}%
\bibitem [{\citenamefont {Balzer}\ \emph {et~al.}(2015)\citenamefont {Balzer},
  \citenamefont {Wolf}, \citenamefont {McCulloch}, \citenamefont {Werner},\
  and\ \citenamefont {Eckstein}}]{balzer2015nonthermal}%
  \BibitemOpen
  \bibfield  {author} {\bibinfo {author} {\bibfnamefont {K.}~\bibnamefont
  {Balzer}}, \bibinfo {author} {\bibfnamefont {F.~A.}\ \bibnamefont {Wolf}},
  \bibinfo {author} {\bibfnamefont {I.~P.}\ \bibnamefont {McCulloch}}, \bibinfo
  {author} {\bibfnamefont {P.}~\bibnamefont {Werner}},\ and\ \bibinfo {author}
  {\bibfnamefont {M.}~\bibnamefont {Eckstein}},\ }\bibfield  {title} {\bibinfo
  {title} {Nonthermal melting of n{\'e}el order in the {H}ubbard model},\
  }\href@noop {} {\bibfield  {journal} {\bibinfo  {journal} {Physical Review
  X}\ }\textbf {\bibinfo {volume} {5}},\ \bibinfo {pages} {031039} (\bibinfo
  {year} {2015})}\BibitemShut {NoStop}%
\bibitem [{\citenamefont {Eckstein}\ and\ \citenamefont
  {Werner}(2013)}]{eckstein2013photoinduced}%
  \BibitemOpen
  \bibfield  {author} {\bibinfo {author} {\bibfnamefont {M.}~\bibnamefont
  {Eckstein}}\ and\ \bibinfo {author} {\bibfnamefont {P.}~\bibnamefont
  {Werner}},\ }\bibfield  {title} {\bibinfo {title} {Photoinduced states in a
  {M}ott insulator},\ }\href@noop {} {\bibfield  {journal} {\bibinfo  {journal}
  {Physical Review Letters}\ }\textbf {\bibinfo {volume} {110}},\ \bibinfo
  {pages} {126401} (\bibinfo {year} {2013})}\BibitemShut {NoStop}%
\bibitem [{\citenamefont {Perfetti}\ \emph {et~al.}(2008)\citenamefont
  {Perfetti}, \citenamefont {Loukakos}, \citenamefont {Lisowski}, \citenamefont
  {Bovensiepen}, \citenamefont {Wolf}, \citenamefont {Berger}, \citenamefont
  {Biermann},\ and\ \citenamefont {Georges}}]{perfetti2008femtosecond}%
  \BibitemOpen
  \bibfield  {author} {\bibinfo {author} {\bibfnamefont {L.}~\bibnamefont
  {Perfetti}}, \bibinfo {author} {\bibfnamefont {P.~A.}\ \bibnamefont
  {Loukakos}}, \bibinfo {author} {\bibfnamefont {M.}~\bibnamefont {Lisowski}},
  \bibinfo {author} {\bibfnamefont {U.}~\bibnamefont {Bovensiepen}}, \bibinfo
  {author} {\bibfnamefont {M.}~\bibnamefont {Wolf}}, \bibinfo {author}
  {\bibfnamefont {H.}~\bibnamefont {Berger}}, \bibinfo {author} {\bibfnamefont
  {S.}~\bibnamefont {Biermann}},\ and\ \bibinfo {author} {\bibfnamefont
  {A.}~\bibnamefont {Georges}},\ }\bibfield  {title} {\bibinfo {title}
  {Femtosecond dynamics of electronic states in the {M}ott insulator
  1{T}-{T}a{S}$_2$ by time resolved photoelectron spectroscopy},\ }\href@noop
  {} {\bibfield  {journal} {\bibinfo  {journal} {New Journal of Physics}\
  }\textbf {\bibinfo {volume} {10}},\ \bibinfo {pages} {053019} (\bibinfo
  {year} {2008})}\BibitemShut {NoStop}%
\bibitem [{\citenamefont {Morrison}\ \emph {et~al.}(2014)\citenamefont
  {Morrison}, \citenamefont {Chatelain}, \citenamefont {Tiwari}, \citenamefont
  {Hendaoui}, \citenamefont {Bruh{\'a}cs}, \citenamefont {Chaker},\ and\
  \citenamefont {Siwick}}]{morrison2014photoinduced}%
  \BibitemOpen
  \bibfield  {author} {\bibinfo {author} {\bibfnamefont {V.~R.}\ \bibnamefont
  {Morrison}}, \bibinfo {author} {\bibfnamefont {R.~P.}\ \bibnamefont
  {Chatelain}}, \bibinfo {author} {\bibfnamefont {K.~L.}\ \bibnamefont
  {Tiwari}}, \bibinfo {author} {\bibfnamefont {A.}~\bibnamefont {Hendaoui}},
  \bibinfo {author} {\bibfnamefont {A.}~\bibnamefont {Bruh{\'a}cs}}, \bibinfo
  {author} {\bibfnamefont {M.}~\bibnamefont {Chaker}},\ and\ \bibinfo {author}
  {\bibfnamefont {B.~J.}\ \bibnamefont {Siwick}},\ }\bibfield  {title}
  {\bibinfo {title} {A photoinduced metal-like phase of monoclinic {V}{O}$_2$
  revealed by ultrafast electron diffraction},\ }\href@noop {} {\bibfield
  {journal} {\bibinfo  {journal} {Science}\ }\textbf {\bibinfo {volume}
  {346}},\ \bibinfo {pages} {445} (\bibinfo {year} {2014})}\BibitemShut
  {NoStop}%
\bibitem [{\citenamefont {Sinn}\ \emph {et~al.}(2016)\citenamefont {Sinn},
  \citenamefont {Kim}, \citenamefont {Kim}, \citenamefont {Lee}, \citenamefont
  {Won}, \citenamefont {Oh}, \citenamefont {Han}, \citenamefont {Chang},
  \citenamefont {Hur}, \citenamefont {Sato} \emph
  {et~al.}}]{sinn2016electronic}%
  \BibitemOpen
  \bibfield  {author} {\bibinfo {author} {\bibfnamefont {S.}~\bibnamefont
  {Sinn}}, \bibinfo {author} {\bibfnamefont {C.~H.}\ \bibnamefont {Kim}},
  \bibinfo {author} {\bibfnamefont {B.~H.}\ \bibnamefont {Kim}}, \bibinfo
  {author} {\bibfnamefont {K.~D.}\ \bibnamefont {Lee}}, \bibinfo {author}
  {\bibfnamefont {C.~J.}\ \bibnamefont {Won}}, \bibinfo {author} {\bibfnamefont
  {J.~S.}\ \bibnamefont {Oh}}, \bibinfo {author} {\bibfnamefont
  {M.}~\bibnamefont {Han}}, \bibinfo {author} {\bibfnamefont {Y.~J.}\
  \bibnamefont {Chang}}, \bibinfo {author} {\bibfnamefont {N.}~\bibnamefont
  {Hur}}, \bibinfo {author} {\bibfnamefont {H.}~\bibnamefont {Sato}}, \emph
  {et~al.},\ }\bibfield  {title} {\bibinfo {title} {Electronic structure of the
  {K}itaev material $\alpha$-{R}u{C}l$_3$ probed by photoemission and inverse
  photoemission spectroscopies},\ }\href@noop {} {\bibfield  {journal}
  {\bibinfo  {journal} {Scientific reports}\ }\textbf {\bibinfo {volume} {6}},\
  \bibinfo {pages} {1} (\bibinfo {year} {2016})}\BibitemShut {NoStop}%
\bibitem [{\citenamefont {Prasankumar}\ \emph {et~al.}(2005)\citenamefont
  {Prasankumar}, \citenamefont {Okamura}, \citenamefont {Imai}, \citenamefont
  {Shimakawa}, \citenamefont {Kubo}, \citenamefont {Trugman}, \citenamefont
  {Taylor},\ and\ \citenamefont {Averitt}}]{prasankumar2005coupled}%
  \BibitemOpen
  \bibfield  {author} {\bibinfo {author} {\bibfnamefont {R.}~\bibnamefont
  {Prasankumar}}, \bibinfo {author} {\bibfnamefont {H.}~\bibnamefont
  {Okamura}}, \bibinfo {author} {\bibfnamefont {H.}~\bibnamefont {Imai}},
  \bibinfo {author} {\bibfnamefont {Y.}~\bibnamefont {Shimakawa}}, \bibinfo
  {author} {\bibfnamefont {Y.}~\bibnamefont {Kubo}}, \bibinfo {author}
  {\bibfnamefont {S.}~\bibnamefont {Trugman}}, \bibinfo {author} {\bibfnamefont
  {A.}~\bibnamefont {Taylor}},\ and\ \bibinfo {author} {\bibfnamefont
  {R.}~\bibnamefont {Averitt}},\ }\bibfield  {title} {\bibinfo {title} {Coupled
  charge-spin dynamics of the magnetoresistive pyrochlore
  {T}l$_2${M}n$_2${O}$_7$ probed using ultrafast midinfrared spectroscopy},\
  }\href@noop {} {\bibfield  {journal} {\bibinfo  {journal} {Physical Review
  Letters}\ }\textbf {\bibinfo {volume} {95}},\ \bibinfo {pages} {267404}
  (\bibinfo {year} {2005})}\BibitemShut {NoStop}%
\bibitem [{\citenamefont {Eckstein}\ and\ \citenamefont
  {Werner}(2014)}]{eckstein2014ultrafast}%
  \BibitemOpen
  \bibfield  {author} {\bibinfo {author} {\bibfnamefont {M.}~\bibnamefont
  {Eckstein}}\ and\ \bibinfo {author} {\bibfnamefont {P.}~\bibnamefont
  {Werner}},\ }\bibfield  {title} {\bibinfo {title} {Ultrafast separation of
  photodoped carriers in {M}ott antiferromagnets},\ }\href@noop {} {\bibfield
  {journal} {\bibinfo  {journal} {Physical Review Letters}\ }\textbf {\bibinfo
  {volume} {113}},\ \bibinfo {pages} {076405} (\bibinfo {year}
  {2014})}\BibitemShut {NoStop}%
\bibitem [{\citenamefont {Shpyrko}\ \emph {et~al.}(2007)\citenamefont
  {Shpyrko}, \citenamefont {Isaacs}, \citenamefont {Logan}, \citenamefont
  {Feng}, \citenamefont {Aeppli}, \citenamefont {Jaramillo}, \citenamefont
  {Kim}, \citenamefont {Rosenbaum}, \citenamefont {Zschack}, \citenamefont
  {Sprung} \emph {et~al.}}]{shpyrko2007direct}%
  \BibitemOpen
  \bibfield  {author} {\bibinfo {author} {\bibfnamefont {O.}~\bibnamefont
  {Shpyrko}}, \bibinfo {author} {\bibfnamefont {E.}~\bibnamefont {Isaacs}},
  \bibinfo {author} {\bibfnamefont {J.}~\bibnamefont {Logan}}, \bibinfo
  {author} {\bibfnamefont {Y.}~\bibnamefont {Feng}}, \bibinfo {author}
  {\bibfnamefont {G.}~\bibnamefont {Aeppli}}, \bibinfo {author} {\bibfnamefont
  {R.}~\bibnamefont {Jaramillo}}, \bibinfo {author} {\bibfnamefont
  {H.}~\bibnamefont {Kim}}, \bibinfo {author} {\bibfnamefont {T.}~\bibnamefont
  {Rosenbaum}}, \bibinfo {author} {\bibfnamefont {P.}~\bibnamefont {Zschack}},
  \bibinfo {author} {\bibfnamefont {M.}~\bibnamefont {Sprung}}, \emph
  {et~al.},\ }\bibfield  {title} {\bibinfo {title} {Direct measurement of
  antiferromagnetic domain fluctuations},\ }\href@noop {} {\bibfield  {journal}
  {\bibinfo  {journal} {Nature}\ }\textbf {\bibinfo {volume} {447}},\ \bibinfo
  {pages} {68} (\bibinfo {year} {2007})}\BibitemShut {NoStop}%
\bibitem [{\citenamefont {Cipelletti}\ \emph {et~al.}(2003)\citenamefont
  {Cipelletti}, \citenamefont {Ramos}, \citenamefont {Manley}, \citenamefont
  {Pitard}, \citenamefont {Weitz}, \citenamefont {Pashkovski},\ and\
  \citenamefont {Johansson}}]{cipelletti2003universal}%
  \BibitemOpen
  \bibfield  {author} {\bibinfo {author} {\bibfnamefont {L.}~\bibnamefont
  {Cipelletti}}, \bibinfo {author} {\bibfnamefont {L.}~\bibnamefont {Ramos}},
  \bibinfo {author} {\bibfnamefont {S.}~\bibnamefont {Manley}}, \bibinfo
  {author} {\bibfnamefont {E.}~\bibnamefont {Pitard}}, \bibinfo {author}
  {\bibfnamefont {D.~A.}\ \bibnamefont {Weitz}}, \bibinfo {author}
  {\bibfnamefont {E.~E.}\ \bibnamefont {Pashkovski}},\ and\ \bibinfo {author}
  {\bibfnamefont {M.}~\bibnamefont {Johansson}},\ }\bibfield  {title} {\bibinfo
  {title} {Universal non-diffusive slow dynamics in aging soft matter},\
  }\href@noop {} {\bibfield  {journal} {\bibinfo  {journal} {Faraday
  discussions}\ }\textbf {\bibinfo {volume} {123}},\ \bibinfo {pages} {237}
  (\bibinfo {year} {2003})}\BibitemShut {NoStop}%
\bibitem [{\citenamefont {Xi}\ \emph {et~al.}(2008)\citenamefont {Xi},
  \citenamefont {Gao}, \citenamefont {Ouyang}, \citenamefont {Shi},\ and\
  \citenamefont {Yang}}]{xi2008slow}%
  \BibitemOpen
  \bibfield  {author} {\bibinfo {author} {\bibfnamefont {H.}~\bibnamefont
  {Xi}}, \bibinfo {author} {\bibfnamefont {K.-Z.}\ \bibnamefont {Gao}},
  \bibinfo {author} {\bibfnamefont {J.}~\bibnamefont {Ouyang}}, \bibinfo
  {author} {\bibfnamefont {Y.}~\bibnamefont {Shi}},\ and\ \bibinfo {author}
  {\bibfnamefont {Y.}~\bibnamefont {Yang}},\ }\bibfield  {title} {\bibinfo
  {title} {Slow magnetization relaxation and reversal in magnetic thin films},\
  }\href@noop {} {\bibfield  {journal} {\bibinfo  {journal} {Journal of
  Physics: Condensed Matter}\ }\textbf {\bibinfo {volume} {20}},\ \bibinfo
  {pages} {295220} (\bibinfo {year} {2008})}\BibitemShut {NoStop}%
\bibitem [{\citenamefont {Labrune}\ \emph {et~al.}(1989)\citenamefont
  {Labrune}, \citenamefont {Andrieu}, \citenamefont {Rio},\ and\ \citenamefont
  {Bernstein}}]{labrune1989time}%
  \BibitemOpen
  \bibfield  {author} {\bibinfo {author} {\bibfnamefont {M.}~\bibnamefont
  {Labrune}}, \bibinfo {author} {\bibfnamefont {S.}~\bibnamefont {Andrieu}},
  \bibinfo {author} {\bibfnamefont {F.}~\bibnamefont {Rio}},\ and\ \bibinfo
  {author} {\bibfnamefont {P.}~\bibnamefont {Bernstein}},\ }\bibfield  {title}
  {\bibinfo {title} {Time dependence of the magnetization process of
  {R}{E}-{T}{M} alloys},\ }\href@noop {} {\bibfield  {journal} {\bibinfo
  {journal} {Journal of Magnetism and Magnetic Materials}\ }\textbf {\bibinfo
  {volume} {80}},\ \bibinfo {pages} {211} (\bibinfo {year} {1989})}\BibitemShut
  {NoStop}%
\bibitem [{\citenamefont {Romanens}\ \emph {et~al.}(2005)\citenamefont
  {Romanens}, \citenamefont {Pizzini}, \citenamefont {Yokaichiya},
  \citenamefont {Bonfim}, \citenamefont {Pennec}, \citenamefont {Camarero},
  \citenamefont {Vogel}, \citenamefont {Sort}, \citenamefont {Garcia},
  \citenamefont {Rodmacq} \emph {et~al.}}]{romanens2005magnetic}%
  \BibitemOpen
  \bibfield  {author} {\bibinfo {author} {\bibfnamefont {F.}~\bibnamefont
  {Romanens}}, \bibinfo {author} {\bibfnamefont {S.}~\bibnamefont {Pizzini}},
  \bibinfo {author} {\bibfnamefont {F.}~\bibnamefont {Yokaichiya}}, \bibinfo
  {author} {\bibfnamefont {M.}~\bibnamefont {Bonfim}}, \bibinfo {author}
  {\bibfnamefont {Y.}~\bibnamefont {Pennec}}, \bibinfo {author} {\bibfnamefont
  {J.}~\bibnamefont {Camarero}}, \bibinfo {author} {\bibfnamefont
  {J.}~\bibnamefont {Vogel}}, \bibinfo {author} {\bibfnamefont
  {J.}~\bibnamefont {Sort}}, \bibinfo {author} {\bibfnamefont {F.}~\bibnamefont
  {Garcia}}, \bibinfo {author} {\bibfnamefont {B.}~\bibnamefont {Rodmacq}},
  \emph {et~al.},\ }\bibfield  {title} {\bibinfo {title} {Magnetic relaxation
  of exchange biased {P}t/{C}o multilayers studied by time-resolved kerr
  microscopy},\ }\href@noop {} {\bibfield  {journal} {\bibinfo  {journal}
  {Physical Review B}\ }\textbf {\bibinfo {volume} {72}},\ \bibinfo {pages}
  {134410} (\bibinfo {year} {2005})}\BibitemShut {NoStop}%
\bibitem [{\citenamefont {Adjanoh}\ \emph {et~al.}(2011)\citenamefont
  {Adjanoh}, \citenamefont {Belhi}, \citenamefont {Vogel}, \citenamefont
  {Ayadi},\ and\ \citenamefont {Abdelmoula}}]{adjanoh2011compressed}%
  \BibitemOpen
  \bibfield  {author} {\bibinfo {author} {\bibfnamefont {A.~A.}\ \bibnamefont
  {Adjanoh}}, \bibinfo {author} {\bibfnamefont {R.}~\bibnamefont {Belhi}},
  \bibinfo {author} {\bibfnamefont {J.}~\bibnamefont {Vogel}}, \bibinfo
  {author} {\bibfnamefont {M.}~\bibnamefont {Ayadi}},\ and\ \bibinfo {author}
  {\bibfnamefont {K.}~\bibnamefont {Abdelmoula}},\ }\bibfield  {title}
  {\bibinfo {title} {Compressed exponential form for disordered domain wall
  motion in ultra-thin {A}u/{C}o/{A}u ferromagnetic films},\ }\href@noop {}
  {\bibfield  {journal} {\bibinfo  {journal} {Journal of Magnetism and Magnetic
  Materials}\ }\textbf {\bibinfo {volume} {323}},\ \bibinfo {pages} {504}
  (\bibinfo {year} {2011})}\BibitemShut {NoStop}%
\bibitem [{\citenamefont {Pommier}\ \emph {et~al.}(1990)\citenamefont
  {Pommier}, \citenamefont {Meyer}, \citenamefont {P{\'e}nissard},
  \citenamefont {Ferr{\'e}}, \citenamefont {Bruno},\ and\ \citenamefont
  {Renard}}]{pommier1990magnetization}%
  \BibitemOpen
  \bibfield  {author} {\bibinfo {author} {\bibfnamefont {J.}~\bibnamefont
  {Pommier}}, \bibinfo {author} {\bibfnamefont {P.}~\bibnamefont {Meyer}},
  \bibinfo {author} {\bibfnamefont {G.}~\bibnamefont {P{\'e}nissard}}, \bibinfo
  {author} {\bibfnamefont {J.}~\bibnamefont {Ferr{\'e}}}, \bibinfo {author}
  {\bibfnamefont {P.}~\bibnamefont {Bruno}},\ and\ \bibinfo {author}
  {\bibfnamefont {D.}~\bibnamefont {Renard}},\ }\bibfield  {title} {\bibinfo
  {title} {Magnetization reversal in ultrathin ferromagnetic films with
  perpendicular anistropy: {D}omain observations},\ }\href@noop {} {\bibfield
  {journal} {\bibinfo  {journal} {Physical Review Letters}\ }\textbf {\bibinfo
  {volume} {65}},\ \bibinfo {pages} {2054} (\bibinfo {year}
  {1990})}\BibitemShut {NoStop}%
\bibitem [{\citenamefont {Cheong}\ \emph {et~al.}(2020)\citenamefont {Cheong},
  \citenamefont {Fiebig}, \citenamefont {Wu}, \citenamefont {Chapon},\ and\
  \citenamefont {Kiryukhin}}]{cheong2020seeing}%
  \BibitemOpen
  \bibfield  {author} {\bibinfo {author} {\bibfnamefont {S.-W.}\ \bibnamefont
  {Cheong}}, \bibinfo {author} {\bibfnamefont {M.}~\bibnamefont {Fiebig}},
  \bibinfo {author} {\bibfnamefont {W.}~\bibnamefont {Wu}}, \bibinfo {author}
  {\bibfnamefont {L.}~\bibnamefont {Chapon}},\ and\ \bibinfo {author}
  {\bibfnamefont {V.}~\bibnamefont {Kiryukhin}},\ }\bibfield  {title} {\bibinfo
  {title} {Seeing is believing: visualization of antiferromagnetic domains},\
  }\href@noop {} {\bibfield  {journal} {\bibinfo  {journal} {npj Quantum
  Materials}\ }\textbf {\bibinfo {volume} {5}},\ \bibinfo {pages} {1} (\bibinfo
  {year} {2020})}\BibitemShut {NoStop}%
\bibitem [{\citenamefont {Chaloupka}\ and\ \citenamefont
  {Khaliullin}(2016)}]{chaloupka2016magnetic}%
  \BibitemOpen
  \bibfield  {author} {\bibinfo {author} {\bibfnamefont {J.}~\bibnamefont
  {Chaloupka}}\ and\ \bibinfo {author} {\bibfnamefont {G.}~\bibnamefont
  {Khaliullin}},\ }\bibfield  {title} {\bibinfo {title} {Magnetic anisotropy in
  the {K}itaev model systems {N}a$_2${I}r{O}$_3$ and $\alpha$-{R}u{C}l$_3$},\
  }\href@noop {} {\bibfield  {journal} {\bibinfo  {journal} {Physical Review
  B}\ }\textbf {\bibinfo {volume} {94}},\ \bibinfo {pages} {064435} (\bibinfo
  {year} {2016})}\BibitemShut {NoStop}%
\bibitem [{\citenamefont {Hedayat}\ \emph {et~al.}(2021)\citenamefont
  {Hedayat}, \citenamefont {Sayers}, \citenamefont {Ceraso}, \citenamefont {van
  Wezel}, \citenamefont {Clark}, \citenamefont {Dallera}, \citenamefont
  {Cerullo}, \citenamefont {Da~Como},\ and\ \citenamefont
  {Carpene}}]{hedayat2021investigation}%
  \BibitemOpen
  \bibfield  {author} {\bibinfo {author} {\bibfnamefont {H.}~\bibnamefont
  {Hedayat}}, \bibinfo {author} {\bibfnamefont {C.~J.}\ \bibnamefont {Sayers}},
  \bibinfo {author} {\bibfnamefont {A.}~\bibnamefont {Ceraso}}, \bibinfo
  {author} {\bibfnamefont {J.}~\bibnamefont {van Wezel}}, \bibinfo {author}
  {\bibfnamefont {S.~R.}\ \bibnamefont {Clark}}, \bibinfo {author}
  {\bibfnamefont {C.}~\bibnamefont {Dallera}}, \bibinfo {author} {\bibfnamefont
  {G.}~\bibnamefont {Cerullo}}, \bibinfo {author} {\bibfnamefont
  {E.}~\bibnamefont {Da~Como}},\ and\ \bibinfo {author} {\bibfnamefont
  {E.}~\bibnamefont {Carpene}},\ }\bibfield  {title} {\bibinfo {title}
  {Investigation of the non-equilibrium state of strongly correlated materials
  by complementary ultrafast spectroscopy techniques},\ }\href@noop {}
  {\bibfield  {journal} {\bibinfo  {journal} {New Journal of Physics}\ }\textbf
  {\bibinfo {volume} {23}},\ \bibinfo {pages} {033025} (\bibinfo {year}
  {2021})}\BibitemShut {NoStop}%
\bibitem [{\citenamefont {Hedayat}\ \emph {et~al.}(2018)\citenamefont
  {Hedayat}, \citenamefont {Bugini}, \citenamefont {Yi}, \citenamefont {Chen},
  \citenamefont {Zhou}, \citenamefont {Cerullo}, \citenamefont {Dallera},\ and\
  \citenamefont {Carpene}}]{hedayat2018surface}%
  \BibitemOpen
  \bibfield  {author} {\bibinfo {author} {\bibfnamefont {H.}~\bibnamefont
  {Hedayat}}, \bibinfo {author} {\bibfnamefont {D.}~\bibnamefont {Bugini}},
  \bibinfo {author} {\bibfnamefont {H.}~\bibnamefont {Yi}}, \bibinfo {author}
  {\bibfnamefont {C.}~\bibnamefont {Chen}}, \bibinfo {author} {\bibfnamefont
  {X.}~\bibnamefont {Zhou}}, \bibinfo {author} {\bibfnamefont {G.}~\bibnamefont
  {Cerullo}}, \bibinfo {author} {\bibfnamefont {C.}~\bibnamefont {Dallera}},\
  and\ \bibinfo {author} {\bibfnamefont {E.}~\bibnamefont {Carpene}},\
  }\bibfield  {title} {\bibinfo {title} {Surface state dynamics of topological
  insulators investigated by femtosecond time-and angle-resolved photoemission
  spectroscopy},\ }\href@noop {} {\bibfield  {journal} {\bibinfo  {journal}
  {Applied Sciences}\ }\textbf {\bibinfo {volume} {8}},\ \bibinfo {pages} {694}
  (\bibinfo {year} {2018})}\BibitemShut {NoStop}%
\bibitem [{\citenamefont {Gerber}\ \emph {et~al.}(2017)\citenamefont {Gerber},
  \citenamefont {Yang}, \citenamefont {Zhu}, \citenamefont {Soifer},
  \citenamefont {Sobota}, \citenamefont {Rebec}, \citenamefont {Lee},
  \citenamefont {Jia}, \citenamefont {Moritz}, \citenamefont {Jia} \emph
  {et~al.}}]{gerber2017femtosecond}%
  \BibitemOpen
  \bibfield  {author} {\bibinfo {author} {\bibfnamefont {S.}~\bibnamefont
  {Gerber}}, \bibinfo {author} {\bibfnamefont {S.-L.}\ \bibnamefont {Yang}},
  \bibinfo {author} {\bibfnamefont {D.}~\bibnamefont {Zhu}}, \bibinfo {author}
  {\bibfnamefont {H.}~\bibnamefont {Soifer}}, \bibinfo {author} {\bibfnamefont
  {J.}~\bibnamefont {Sobota}}, \bibinfo {author} {\bibfnamefont
  {S.}~\bibnamefont {Rebec}}, \bibinfo {author} {\bibfnamefont
  {J.}~\bibnamefont {Lee}}, \bibinfo {author} {\bibfnamefont {T.}~\bibnamefont
  {Jia}}, \bibinfo {author} {\bibfnamefont {B.}~\bibnamefont {Moritz}},
  \bibinfo {author} {\bibfnamefont {C.}~\bibnamefont {Jia}}, \emph {et~al.},\
  }\bibfield  {title} {\bibinfo {title} {Femtosecond electron-phonon lock-in by
  photoemission and x-ray free-electron laser},\ }\href@noop {} {\bibfield
  {journal} {\bibinfo  {journal} {Science}\ }\textbf {\bibinfo {volume}
  {357}},\ \bibinfo {pages} {71} (\bibinfo {year} {2017})}\BibitemShut
  {NoStop}%
\bibitem [{\citenamefont {Versteeg}\ \emph {et~al.}(2018)\citenamefont
  {Versteeg}, \citenamefont {Zhu}, \citenamefont {Padmanabhan}, \citenamefont
  {Boguschewski}, \citenamefont {German}, \citenamefont {Goedecke},
  \citenamefont {Becker},\ and\ \citenamefont
  {Van~Loosdrecht}}]{versteeg2018tunable}%
  \BibitemOpen
  \bibfield  {author} {\bibinfo {author} {\bibfnamefont {R.}~\bibnamefont
  {Versteeg}}, \bibinfo {author} {\bibfnamefont {J.}~\bibnamefont {Zhu}},
  \bibinfo {author} {\bibfnamefont {P.}~\bibnamefont {Padmanabhan}}, \bibinfo
  {author} {\bibfnamefont {C.}~\bibnamefont {Boguschewski}}, \bibinfo {author}
  {\bibfnamefont {R.}~\bibnamefont {German}}, \bibinfo {author} {\bibfnamefont
  {M.}~\bibnamefont {Goedecke}}, \bibinfo {author} {\bibfnamefont
  {P.}~\bibnamefont {Becker}},\ and\ \bibinfo {author} {\bibfnamefont
  {P.}~\bibnamefont {Van~Loosdrecht}},\ }\bibfield  {title} {\bibinfo {title}
  {A tunable time-resolved spontaneous raman spectroscopy setup for probing
  ultrafast collective excitation and quasiparticle dynamics in quantum
  materials},\ }\href@noop {} {\bibfield  {journal} {\bibinfo  {journal}
  {Structural Dynamics}\ }\textbf {\bibinfo {volume} {5}},\ \bibinfo {pages}
  {044301} (\bibinfo {year} {2018})}\BibitemShut {NoStop}%
\end{thebibliography}%



\end{document}


\pagenumbering{arabic}

\title{\fontsize{15}{19}\selectfont Supplementary information for ""}

\author{Julian Wagner}

\author{Anuja Sahasrabudhe}
\affiliation{Universität zu Köln, II. Physikalisches Institut, Zülpicher Straße 77, Köln D-50937, Germany}

\author{Rolf Versteeg}

\affiliation{Laboratoire de Spectroscopie Ultrarapide and Lausanne Centre for Ultrafast Science (LACUS),\\ ISIC-FSB, \'Ecole Polytechnique F\'ed\'erale de Lausanne, CH-1015 Lausanne, Switzerland}

\author{Lena Wysocki}
\affiliation{Universität zu Köln, II. Physikalisches Institut, Zülpicher Straße 77, Köln D-50937, Germany} 

\author{Zhe Wang}
\affiliation{Technische Universität Dortmund, Fakultät Physik August-Schmidt-Str. 4, Dortmund D-44227, Germany}

\author{V.~Tsurkan}
\affiliation{University of Augsburg, 86159 Augsburg, Germany}
\affiliation{ Institute of Applied Physics, MD 2028, Chisinau, Republic of Moldova}

\author{A.~Loidl}
\affiliation{University of Augsburg, 86159 Augsburg, Germany}

\author{D.~I.~Khomskii}

\author{Hamoon Hedayat}
\email{hedayat@ph2.uni-koeln.de}
\author{Paul H.M. van Loosdrecht}
\email{pvl@ph2.uni-koeln.de}
\affiliation{Universität zu Köln, II. Physikalisches Institut, Zülpicher Straße 77, Köln D-50937, Germany}

\maketitle

\section{Sample growth, characterization and orientation}

High-quality $\alpha$-RuCl$_3$ crystals were prepared by vacuum sublimation \cite{PhysRevB.101.140410}. The different samples of the same batch were characterized by SQUID magnetometry, showing a sharp transition at around $T_N\approx 7$K in zero applied field corresponding to the $ABC$ stacking order \cite{PhysRevB.103.174413,PhysRevB.93.134423}. No additional magnetic transitions above $T_N$ are observed, which have been related to a different stacking order $ABAB$ with a two-layer periodicity or strain-introduced stacking faults due to extensive handling or deformation of the crystals. This bulk technique can only provide the first indication of sample quality for an optics study. Cleaving or polishing introduces strain. In this regard, we refrained from any sample treatment and used an as-grown $\alpha$-RuCl$_3$ sample. The temperature dependence of the equilibrium MO response shown in Fig. ~\ref{FigS1} shows a clear phase transition at $T_N\approx 7.3$K, confirming good sample quality (see Sec.~\ref{Temp}).\\
The different $\alpha$-RuCl$_3$ samples have been oriented via a standard x-ray Laue-diffractometer at room temperature (not shown).\\

\def\bibsection{\section*{~\refname}} 
\bibliography{bibliography}